\documentclass[12pt,showpacs,preprintnumbers,preprint,amsmath,amssymb,aps,superscriptaddress,letterpaper,nofootinbib,longbibliography]{revtex4-1}
\usepackage[colorlinks=true,citecolor=blue,linkcolor=magenta]{hyperref}

\usepackage{amsfonts}
\usepackage{rotating}
\usepackage{epsfig,amssymb,euscript}
\usepackage{type1cm}
\usepackage{color}
\usepackage{xr}
\newcommand{\nn}{\nonumber \\}
\newcommand{\reef}[1]{(\ref{#1})}
\def\cO{{\cal O}}

\def\be{\begin{equation}}
\def\ee{\end{equation}}
\def\bea{\begin{eqnarray}}
\def\eea{\end{eqnarray}}

\begin{document}
\title{Universal far-from-equilibrium Dynamics of a \\ Holographic Superconductor}
\author{Julian Sonner\footnote{after 1 July 2015: D\'epartement de Physique Th\'eorique, Universit\'e de Gen\`eve, 1211 Gen\`eve 4, Switzerland }}
\email{sonner@mit.edu}
\affiliation{CTP, Laboratory for Nuclear Science, Massachusetts Institute of Technology, Cambridge, 77 Massachusetts Avenue,  MA 02139, U.S.A.}
\author{Adolfo del Campo}
\affiliation{Department of Physics, University of Massachusetts, Boston, MA 02125, USA}
\affiliation{Theoretical Division, Los Alamos National Laboratory, Los Alamos, NM 87545, USA}
\affiliation{Center for Nonlinear Studies, Los Alamos National Laboratory, 
Los Alamos, NM 87545, USA}
\author{Wojciech H. Zurek}
\affiliation{Theoretical Division, Los Alamos National Laboratory, Los Alamos, NM 87545, USA}

\preprint{MIT-CTP 4553}
\preprint{LA-UR-14-24054}


\begin{abstract} 
Symmetry breaking phase transitions are an example of non-equilibrium processes that require real time treatment, a major challenge in strongly coupled systems without long-lived quasiparticles. Holographic duality provides such an approach by mapping strongly coupled field theories in D dimensions into weakly coupled quantum gravity in D+1 anti-de Sitter spacetime. Here, we use holographic duality to study formation of topological defects -- winding numbers -- in the course of a superconducting transition in a strongly coupled theory in a 1D ring. When the system undergoes the transition on a given quench time, the condensate builds up with a delay that can be deduced using the Kibble-Zurek mechanism from the quench time and the universality class of the theory, as determined from the quasinormal mode spectrum of the dual model. Typical winding numbers deposited in the ring exhibit a universal fractional power law dependence on the quench time, also  predicted by the Kibble-Zurek Mechanism.  
\end{abstract}


\maketitle

\hypersetup{
    linkcolor=black,          
} 
\pagebreak
\hypersetup{
    linkcolor=magenta,          
} 
\section{Introduction}
Non-equilibrium quantum phenomena are of wide importance across several disciplines of physics. Despite their fundamental relevance, few widely applicable principles are known for field theories far from equilibrium \cite{Dziarmaga10,Polkovnikov11,delCZ14}. Gauge-gravity duality is a powerful tool in this respect, as it provides a well-defined first-principles framework to study strongly-coupled field theories in the non-equilibrium setting. These theories are strongly correlated in the sense that there exists no weakly-coupled quasi-particle picture upon which one could base a perturbative treatment. This makes understanding their dynamics, even near equilibrium, an extremely challenging problem. Consequently progress has been confined mostly to cases in which the dynamics is integrable \cite{calabrese2006time,calabrese2011quantum,fioretto2010quantum,PhysRevLett.109.175301}. 

Holography \cite{Maldacena:1997re} is a powerful tool enabling us to explicitly analyze  such theories without having to rely on integrability: One maps the quantum-field theory of interest to a dual gravity problem in asymptotically Anti de Sitter (AdS) space-time, which can be solved in great detail. This duality can reveal significant new insights, for example it provides new examples of interaction driven localization transitions \cite{donos2013interaction}, as well as modelling finite-temperature transport near the superfluid/insulator critical point \cite{sachdev2014transport}. 

Previous work on dynamics of holographic superfluids analyzed the condensation process by perturbing an uncondensed initial state below criticality \cite{Murata:2010dx}, as well as the non-equilibrium phase diagram resulting from sudden quenches of the superfluid phase  \cite{Bhaseen:2012gg}. Both of these studies focus on spatially homogeneous configurations, as do \cite{Basu:2012gg,Basu:2011ft}, which present scaling results for the timescale of the breakdown of adiabaticity during finite-rate quenches. Subsequent work, relaxing the constraint of spatial homogeneity, explored vortex turbulence of holographic superfluids \cite{Adams:2012pj} (see also \cite{Bai:2014poa} for recent work on anisotropic quenches). This article extends the framework in yet another direction, namely into the regime of non equilibrium dynamics across phase transitions. We analyze the breaking of a continuous $U(1)$ symmetry, by studying the evolution of black holes in AdS, giving a dual description of superconductors. Here we explore a canonical non-equilibrium paradigm associated with broken symmetries, leading to universal scaling results in the formation of defects when a critical point is crossed at a finite rate \cite{kibble1976topology,zurek1985cosmological}. In this context, inhomogeneous configurations of the order parameter are known to be crucial  during  the formation of topological defects. 

\textbf{Note:} Independently, Chesler, Liu and Garcia-Garcia \cite{Chesler:2014gya} have explored the Kibble-Zurek scaling
via the dual black hole quasi-normal mode spectrum and numerical analysis in two spatial
dimensions.

\section{Results}
\subsection{Winding  number generation}
Phase transitions were traditionally studied as equilibrium phenomena. Thus, in the broken symmetry phase, the whole system was assumed to make the same selection of the broken symmetry vacuum. However, as noted in the cosmological context, where a sequence of symmetry breaking phase transitions is thought to have resulted in the familiar fundamental forces, rapid cooling of the post Big Bang Universe combined with relativistic causality makes it impossible to coordinate symmetry breaking outside of the Hubble horizon. As a consequence, distinct domains of the Universe will choose broken symmetry vacua on their own. The resulting mosaic of broken symmetry vacua will -- in the course of the subsequent phase ordering -- attempt to smooth out. As Kibble pointed out  \cite{kibble1976topology}, 
these disparate choices can crystalize into topological defects that may have significant consequences for the subsequent evolution of the Universe.

As noted by one of us \cite{zurek1985cosmological}, systems undergoing second order phase transitions cannot ever follow a sequence of instantaneous equilibria. This is because of the critical slowing down in the vicinity of the critical point: the relaxation time $\tau$ of the order parameter diverges as a function of the dimensionless distance $\epsilon$ from e.g., the critical temperature $T_{\rm c}$:
\be\label{eq:epsilon1}
\epsilon= \frac {T-T_{\rm c}} {T_{\rm c}},
\ee
where $T$ denotes the instantaneous temperature of the system.
This implies that the ``reflexes'' of the system are characterized by the universal power-law,
\be\label{eq:tau}
\tau(\epsilon) \sim \tau_0 |\epsilon|^{-z\nu},
\ee
where $z$ and $\nu$ are the dynamic and correlation length critical exponents, respectively and $\tau_0$ is a microscopic parameter.
An arbitrary cooling ramp as a function of time, $t$, can be linearized around the critical point 
\be\label{eq:epsilon2}
\epsilon= -\frac t {\tau_{\rm Q}}.
\ee
As a result of critical slowing down,  the system loses the ability to adjust to the change even when it happens slowly, on any finite quench timescale $\tau_{\rm Q}$.
The instant $\hat t$ when the system can no longer keep up  with the change of $\epsilon$ happens when its relaxation time becomes comparable with $ \epsilon / {\dot \epsilon}$, the rate of change of $\epsilon$. This leads \cite{zurek1985cosmological} to the equation: 
\be\label{eq:hat_t1}
\tau(\epsilon(\hat t)) = \epsilon / {\dot \epsilon} = \hat t.
\ee
Using the critical slowing down scaling relation, Eq. \reef{eq:tau}, one obtains:
\be\label{eq:hat_t2}
\hat t = \left(\tau_0 \tau_{\rm Q}^{\nu z}\right)^{\frac 1 {1+\nu z}}.
\ee
This time scale allows us to split the crossing of the phase transition into a sequence of three stages  in which the dynamics is first adiabatic ($t<-\hat t$ ), 
then  effectively impulse during the interval $(-\hat{t}, \hat{t} )$, and finally adiabatic again deep in the broken-symmetry side of the transition ($t>\hat t$ ).
The value of $\hat \epsilon$ corresponding to the time scale separating the frozen and adiabatic stages,
\be\label{eq:hat_epsilon}
\hat \epsilon = \left(\frac {\tau_0} {\tau_{\rm Q}} \right)^{\frac 1 {1+\nu z}},
\ee
enters into the sonic horizon estimate, the analog of the causal horizon in the early universe. As a result, the characteristic size $\hat \xi$ of the domains that can coordinate the choice of broken symmetries exhibits a universal power-law dependence \cite{zurek1985cosmological} on the quench time: 
\be\label{eq:hat_xi}
\hat \xi = \xi_0 \hat \epsilon^{-\nu} = \xi_0 \left(\frac {\tau_{\rm Q}} {\tau_0} \right)^{\frac {\nu} {1+\nu z}},
\ee
where $\xi_0$ is a microscopic parameter.
The density of topological defects can then be estimated by recognizing that a $\hat \xi$-sized fragment of defect can be expected within a volume $\hat \xi$-sized domain, the sonic horizon size. Here the velocity of the relevant sound assumes the role of the speed of light in the relativistic cosmological setting. This reasoning is expected to yield correct scaling of the density of defects, but only an order of magnitude estimate of the prefactor. The scenario just described is often referred to as the `Kibble-Zurek mechanism' and we will refer to it throughout this paper as `KZM'.

Here we consider the setting where the phase transition happens in a ring of circumference $C$. In view of our above discussion one can expect that the broken symmetry will be chosen independently in sections of size $\hat \xi$, so there will be $C / \hat \xi$ fragments of the ring that select broken symmetry independently \cite{zurek1985cosmological}. Consequently, phase mismatch resulting from a random walk of phase with $C / \hat \xi$ steps needed to circumnavigate $C$ is expected to lead to:
$\Delta \Theta \approx \sqrt{ \frac C {\hat \xi}}$.
The net ``phase distance''  $\Delta \Theta$ will then have to settle to a multiple of $2\pi$, defining the winding number, $W$, as
\be\label{eq:Winding}
W=\frac {\Delta \Theta} {2 \pi} \approx \frac 1 {2 \pi} \sqrt{ \frac C {\hat \xi}}.
\ee 
It follows that the dispersion of the values of $W$ will scale with the quench rate as \cite{sabbatini2011phase,das2012winding}:
\be\label{eq.Wscale}
\sigma(W) = \frac{1}{2}\sqrt{\frac{C}{3 \xi_0}} \,\left( \frac{\tau_{\rm Q}}{\tau_0}\right)^{-\frac{\nu}{2(1 + z\nu)}}.
\ee
This universal power-law is expected to hold whenever $ C/\hat{\xi}\gg1$,  away from the onset of adiabatic dynamics. 
The prefactor has to be taken with the proverbial ``grain of salt''.  As seen in previous numerical experiments, such KZM calculations yield correct scalings, but tend to overestimate the density of defects \cite{laguna1997density,antunes1999vortex,DeChiara10,sabbatini2011phase} as well as the typical winding numbers \cite{das2012winding,Nigmatullin11}. 

\subsection{A superconducting ring in holography}
\begin{figure}[t!]
\begin{center}
\includegraphics[width=0.7\textwidth]{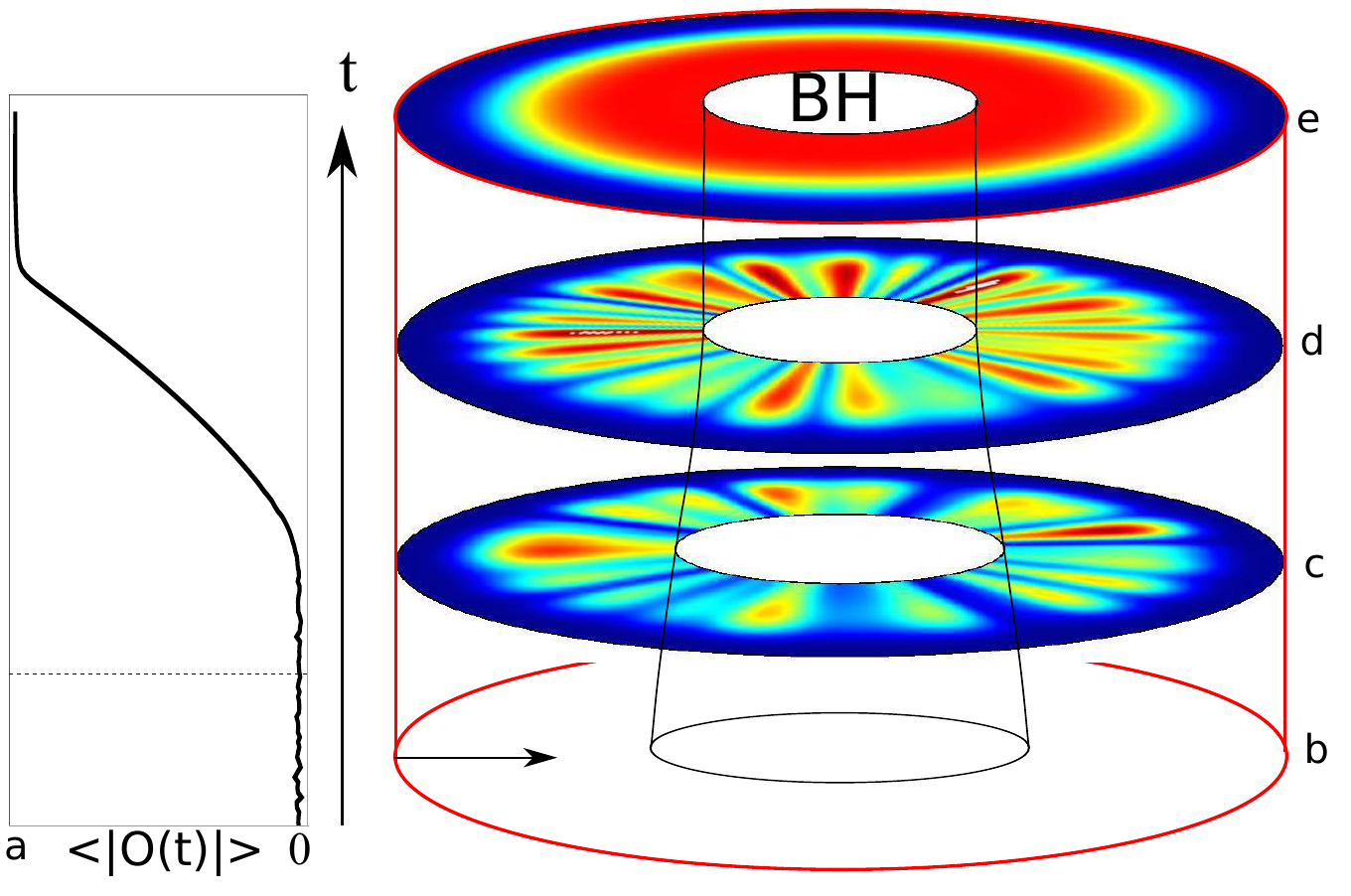}
\begin{picture}(0.1,0.25)(0,0)
\put(-75,14){\makebox(0,0){$r$}}
\put(-88.5,19){\makebox(0,0){$t_{\rm c}$}}
\end{picture}
\caption{ {\bf Schematic of the holographic mapping.} 
The dual space-time is a cylinder, with time running vertically upwards. The field theory lives on the mantle, also called the boundary and shown here in red, an infinite proper distance from the bulk.  In the interior resides a black hole (BH) with temperature $C/2\pi z_{\rm h}$, which is also the temperature of the dual field theory.  Cooling corresponds to the change of the black hole radius $z_{\rm h}$. For convenience all equations in this paper use the coordinate $z \in (0,z_{\rm h}]$, which is related to the more conventional AdS radial coordinate via $\varrho= L^2/z \in [ \varrho_{\rm h}, \infty)$. The radial coordinate $r$ is a compactified version of $\varrho$, which puts the boundary at a finite distance. Panel \textsf{a} schematically shows the order parameter as function of time, spatially averaged over $C$. The four time slices, \textsf{b} - \textsf{e}, indicate the condensation process as it happens throughout the bulk. The density profile shown in each slice schematically illustrates the behavior of the bulk field $\psi$ dual to the order parameter in the boundary theory with blue indicating a vanishing density of $|\psi|^2$ and red a high density of $|\psi|^2$. \textsf{b}: At the beginning there is no condensate, but the temperature is starting to be lowered through $T_{\rm c}$, which is reached at time $t_{\rm c}$, as indicated in \textsf{a}. \textsf{c}: Sometime after crossing $T_{\rm c}$ a small, spatially inhomogeneous condensate starts appearing. \textsf{d}: The condensate grows, amplifying the original inhomogeneities into macroscopic phase domains. \textsf{e}: Eventually the condensate settles down to its equilibrium configuration with a given winding number frozen in. \label{fig.AdS/CFT}}
\end{center}\end{figure}
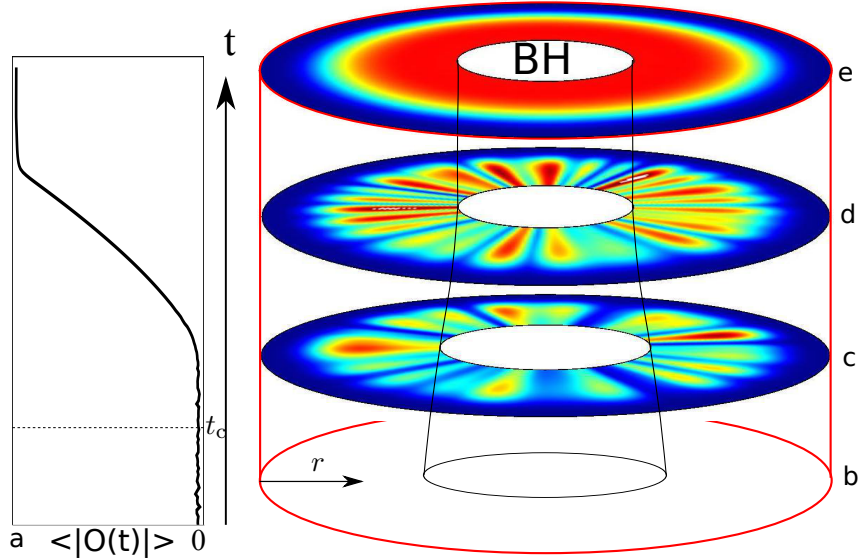
We study a one-dimensional superconducting ring of circumference $C$. The quantity of interest is the winding number
\be
W = \oint_C \frac{d \Theta(\phi)}{2\pi}\,\quad\in\quad\mathbb{Z},
\ee
where $\phi$ is the angle along the ring.
It follows from gauge invariance that in equilibrium we have $\oint \left(d\Theta - A  \right)=0$ so that a non-vanishing winding number is accompanied by a non-vanishing line integral of the vector potential around the loop. It should be noted that $W$ as defined above is gauge invariant under single-valued gauge transformations.

We consider the so-called `bottom-up' holographic superconductor model \cite{Hartnoll:2008vx,Hartnoll:2008kx} in three-dimensional AdS \cite{Ren:2010ha}, denoted from here on AdS$_3$ and work in the probe limit. This amounts to neglecting the effect of the charged components of the system on the neutral ones \cite{Hartnoll:2008vx,Adams:2012pj}.  From the gravity point of view this corresponds to neglecting the bulk gravitational backreaction of the charged scalar and the Maxwell field, but keeping the backreaction of the scalar and Maxwell field on each other. This means that we consistently study the dynamics of a Maxwell field $A$ coupled to a scalar field $\psi$  on a fixed gravitational background. We choose the scalar field to have vanishing mass, which means that it is dual to a classically marginal operator in the boundary field theory. We do not expect our results to differ qualitatively for other choices of the mass, as long as it remains in the range corresponding to a marginal or relevant operator (we refer here to the renormalization group, `RG', scaling dimension in the ultra-violet, `UV'). The details of our bulk action and equations of motion can be found in the Methods section. Maxwell theory without the charged scalar in AdS$_3$ and its dual field theory have been studied previously in \cite{Marolf:2006nd,Jensen:2010em,Faulkner:2012gt}, taking advantage of the fact that one may conveniently dualize the bulk vector field to a bulk scalar. A different holographic approach to superconducting rings, using probe branes, was developed in \cite{Khlebnikov:2012yd,Khlebnikov:2012ny}.

Studying the field theory at finite temperature corresponds to studying the bulk system in the background of a black hole, in our case the three-dimensional black hole (the BTZ black hole, after Ba\~nados, Teitelboim and Zanelli, \cite{Banados:1992wn}), which is characterized by the parameter $z_{\rm h}$, its horizon radius. For more details see Fig. \ref{fig.AdS/CFT}. With this data we have the Hawking temperature
\be\label{eq.Thawk}
T_{\rm H} = \frac{1}{2\pi z_{\rm h}}.
\ee
\begin{figure}[t!]
\begin{center}
\includegraphics[width=\textwidth]{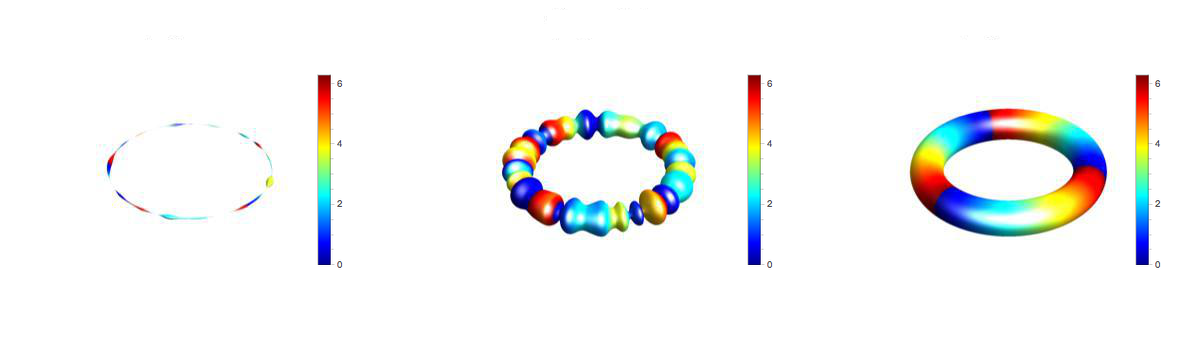}\\
\begin{picture}(0.1,0.25)(0,0)
\put(-80,15){\makebox(0,0){\textsf{a}}}
\put(-25,15){\makebox(0,0){\textsf{b}}}
\put(32,15){\makebox(0,0){\textsf{c}}}
\put(-55,55){\makebox(0,0){$ \mathsf{t/\tau_Q =0.72}$}}
\put(0,55){\makebox(0,0){$ \mathsf{t/\tau_Q =0.82}$}}
\put(55,55){\makebox(0,0){$ \mathsf{t/\tau_Q =1.72}$}}
\put(-42,45){\makebox(0,0){$\Theta$}}
\put(15,45){\makebox(0,0){$\Theta$}}
\put(69,45){\makebox(0,0){$\Theta$}}
\end{picture}
\includegraphics[width=\textwidth]{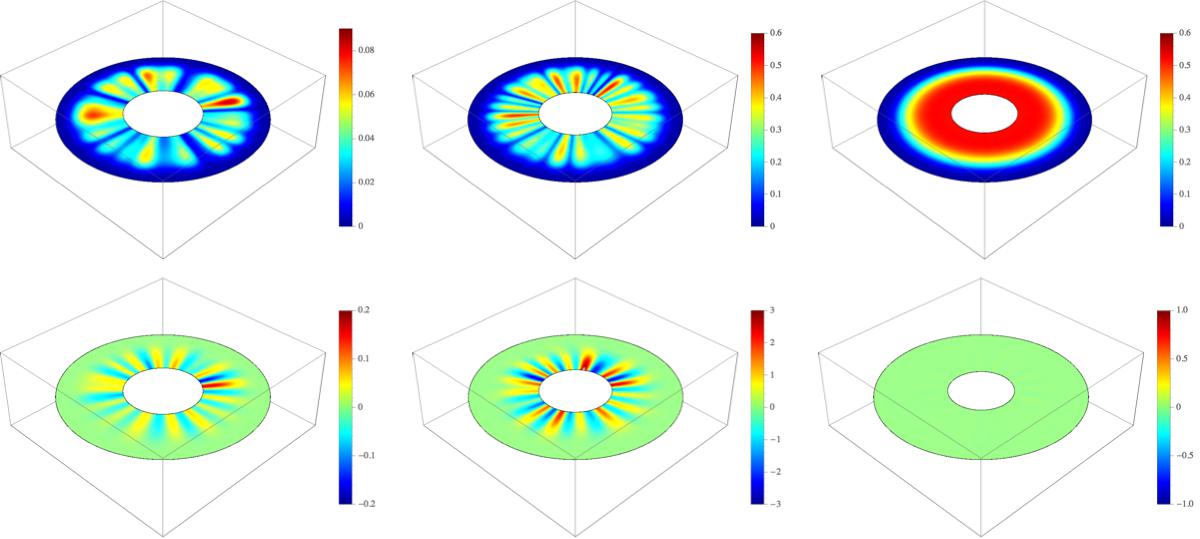}
\begin{picture}(0.1,0.25)(0,0)
\put(12.5,66){\makebox(0,0){$r$}}
\linethickness{0.5pt}
\put(-2.5,66){\vector(1,0){12}}
\put(-44,66){\makebox(0,0){$r$}}
\linethickness{0.5pt}
\put(-59,66){\vector(1,0){12}}
\put(68.5,66){\makebox(0,0){$r$}}
\linethickness{0.5pt}
\put(53,66){\vector(1,0){12}}
\put(12.5,28){\makebox(0,0){$r$}}
\linethickness{0.5pt}
\put(-2.5,28){\vector(1,0){12}}
\put(-44,28){\makebox(0,0){$r$}}
\linethickness{0.5pt}
\put(-59,28){\vector(1,0){12}}
\put(68.5,28){\makebox(0,0){$r$}}
\linethickness{0.5pt}
\put(53,28){\vector(1,0){12}}
\put(-80,48){\makebox(0,0){\textsf{d}}}
\put(-25,48){\makebox(0,0){\textsf{e}}}
\put(32,48){\makebox(0,0){\textsf{f}}}
\put(-80,8){\makebox(0,0){\textsf{g}}}
\put(-25,8){\makebox(0,0){\textsf{h}}}
\put(32,8){\makebox(0,0){\textsf{i}}}
\put(-38,80){\makebox(0,0){$|\psi|^2$}}
\put(18,80){\makebox(0,0){$|\psi|^2$}}
\put(74,80){\makebox(0,0){$|\psi|^2$}}
\put(-38,40){\makebox(0,0){$B$}}
\put(18,40){\makebox(0,0){$B$}}
\put(74,40){\makebox(0,0){$B$}}
\end{picture}
\caption{{\bf Winding up a superconductor following a temperature quench.} Example condensation process with $\tau_{\rm Q} = 12$ leading to a $W = 3$ configuration, shown in three stages.
We choose a gauge that makes the final phase of the order parameter linear $\Theta(\phi) = W\phi$. 
The top row (panels \textsf{a}, \textsf{b}, \textsf{c}) shows the field-theory condensation process; the local magnitude of the condensate is represented as the radius of the torus, while the local phase is encoded as the color, ranging from [$0:$ red] to [$2\pi:$ blue]. The middle row (\textsf{d} - \textsf{f}) shows the magnitude $|\psi|^2$ of the bulk field dual to the order parameter ${\cal O}$ while the last row (\textsf{g} - \textsf{i}) shows the magnetic field $B$ in the bulk. The bulk magnetic fields disappear at late times and the winding around the boundary circle is accompanied by winding number around the horizon of equal magnitude. 
\label{fig.Summary}}
\end{center}\end{figure}
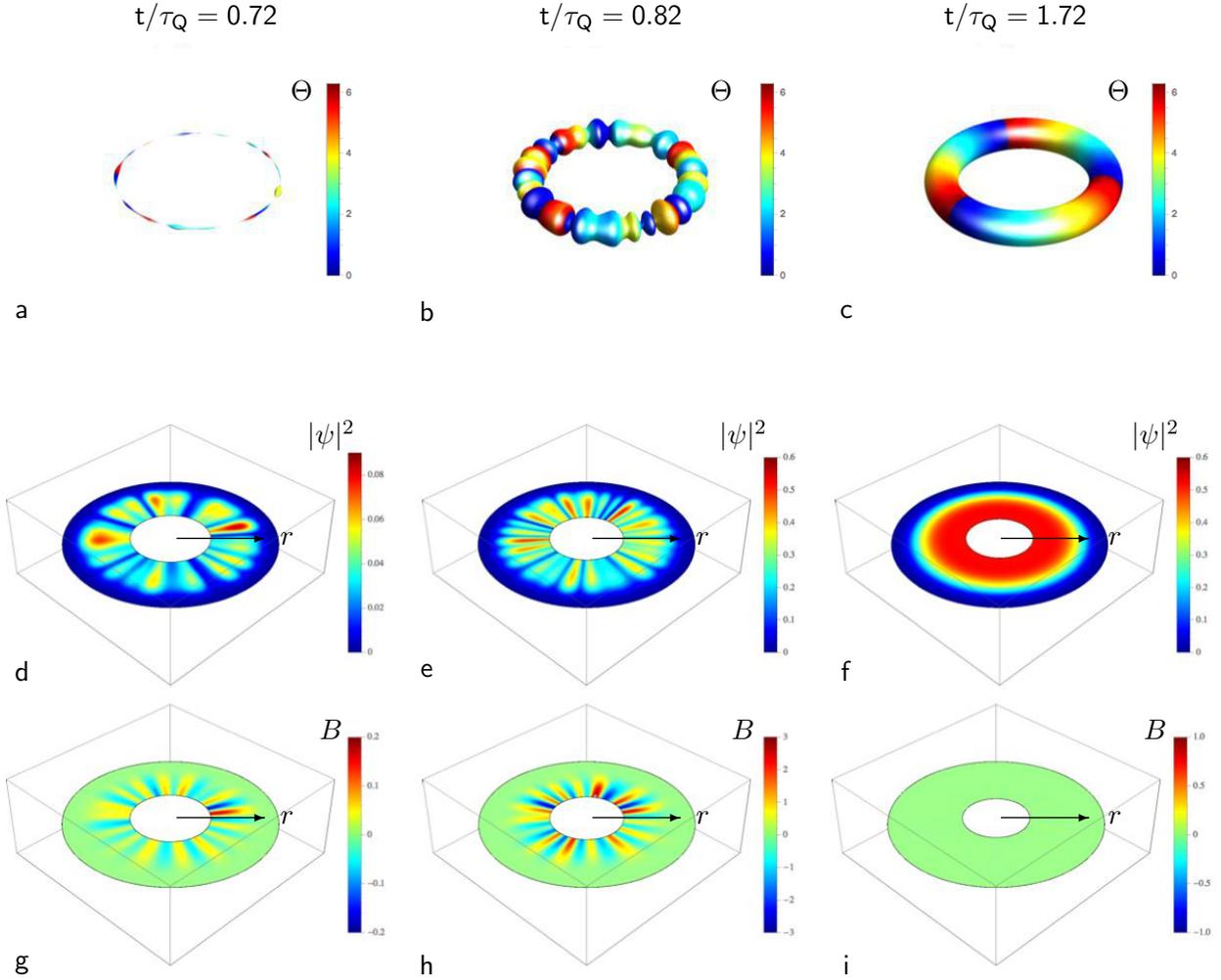
Via the AdS/CFT correspondence (CFT stands for conformal field theory) this is directly translated into the temperature of the dual field theory \cite{Witten:1998zw}. This theory is therefore studied in a thermal ensemble at temperature $T_{\rm H}$, which is an external parameter in our dynamics. In order to fully specify the ensemble (in the equilibrium case), as well as the dynamics we must augment the equations of motion with suitable boundary conditions at the UV boundary $z=0$. We impose in addition to the finite temperature $T$ a fixed charge density $\rho$ corresponding to Neumann boundary conditions on the bulk gauge field. We give Dirichlet boundary conditions for the scalar field. From this it follows that the field theory is studied in the absence of a source for the order parameter density. This condition ensures that any symmetry breaking occurring will be spontaneous. In order to compare our full non-equilibrium results to the  prediction of KZM, we first need to determine the universality class of the field theory just defined, as well as the microscopic parameters $\tau_0$ and $\xi_0$. As is usual in holography, this boils down to computing the so-called quasinormal modes of the bulk black hole \cite{Horowitz:1999jd}.

\subsection{Near equilibrium universality}\label{sec.UniversalityClass}

In order to determine the critical exponents $z,\nu$, as well as the microscopic parameters $\tau_0$ and $\xi_0$, it is sufficient to study the holographic superconductor near equilibrium. We will study the non-compact case, since the finite size of the ring in our simulations does not affect the results for local correlations, so long as the healing length $\xi$ is much smaller than the circumference. The non-compact spatial boundary direction is denoted $x$ here. Since $\xi$ formally diverges near $T_{\rm c}$, this assumption will break down for extremely slow quenches, but we have not found that our simulations entered this regime for the range of $\tau_{\rm Q}$ under study.

\begin{figure}[h!]
\begin{center}
\textsf{a}\includegraphics[width=0.3\textwidth]{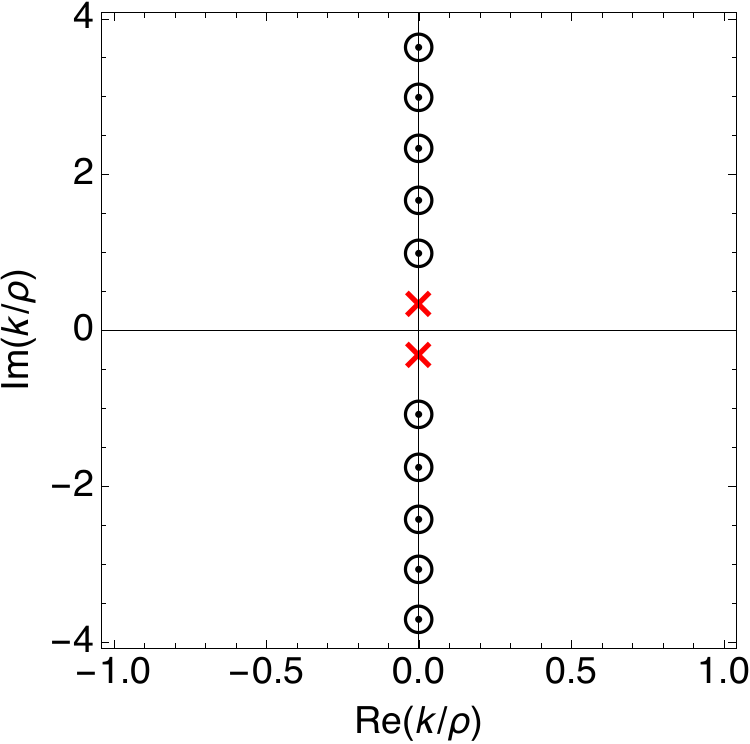}\hskip6em \textsf{b}\includegraphics[width=0.3\textwidth]{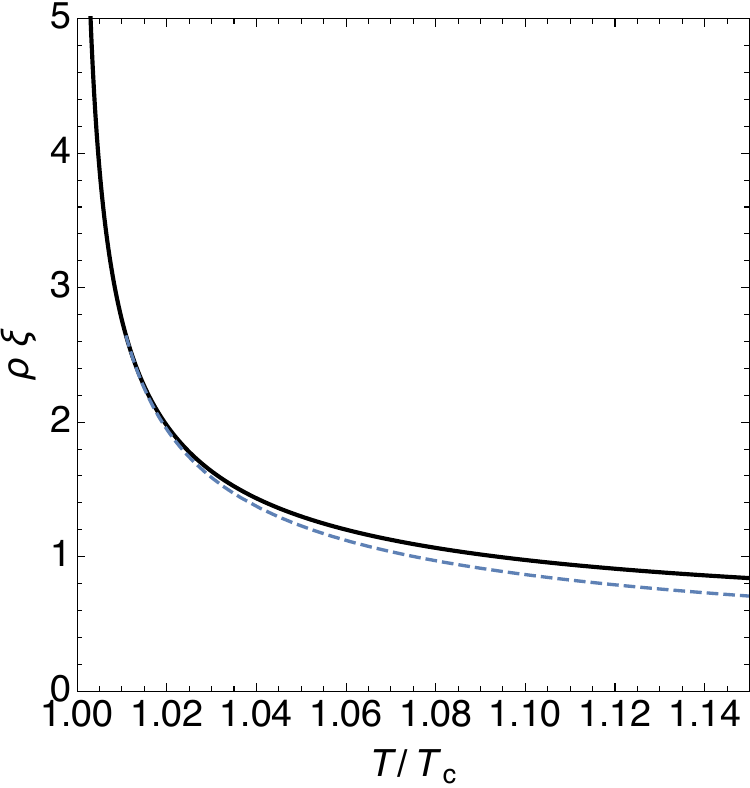}\vskip2em
\textsf{c} \includegraphics[width=0.32\textwidth]{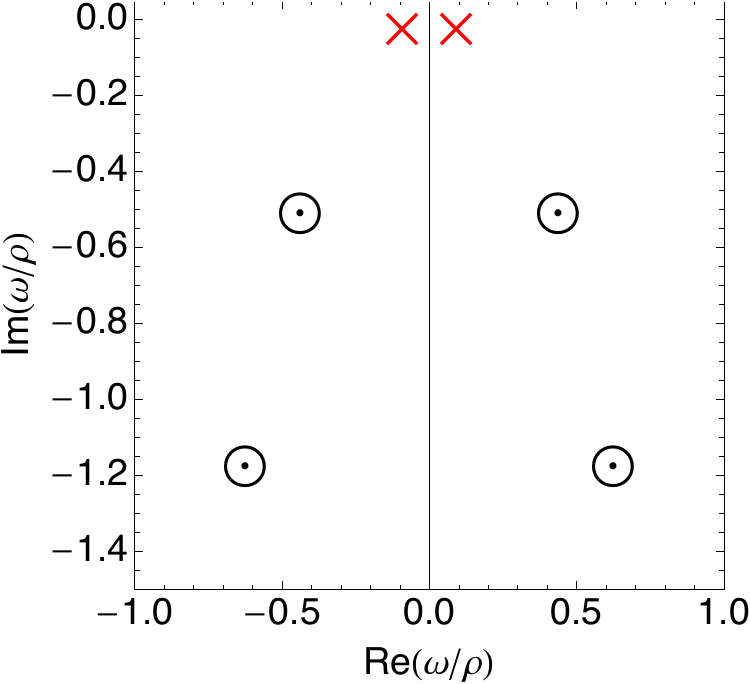}\hskip6em \textsf{d}\includegraphics[width=0.32\textwidth]{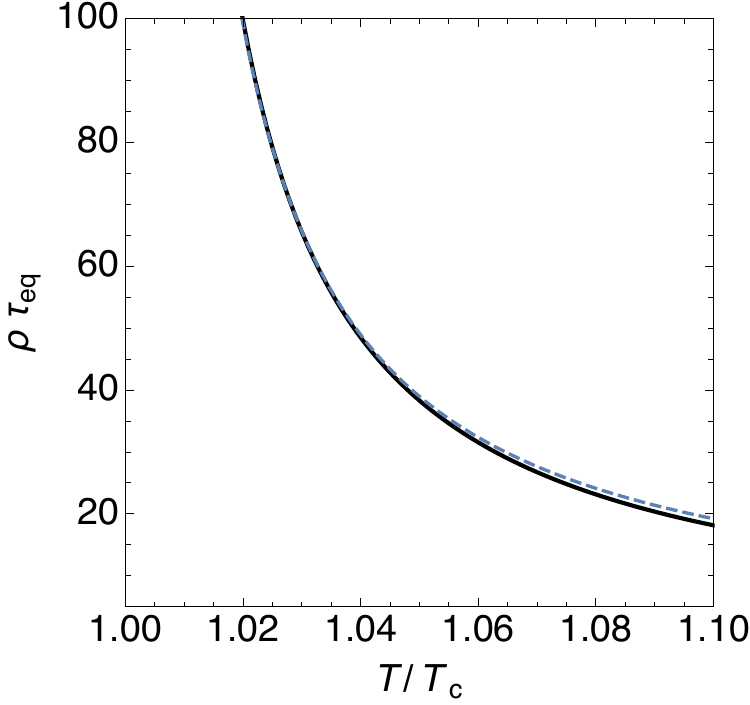}
\caption{{\bf Susceptibilities in the normal phase.} Panels \textsf{a} \& \textsf{c}: Poles in correlation functions at $T/T_{\rm c}=1.1$, as deduced from studying the quasinormal modes of the bulk black hole. These modes are linear fluctuations with dissipative boundary conditions at the horizon. The leading poles nearest the real axis, marked by red crosses, determine the behavior of correlations at large distances (\textsf{a}) and long times (\textsf{c}). Momentum $k$ and frequency $\omega$ are given in units of charge density $\rho$, while temperature $T$ is given in units of the critical temperature $T_{\rm c}$. Panels \textsf{b} \& \textsf{d}: divergence of correlation length (\textsf{b}) and critical slowing down (\textsf{d}), as determined from the corresponding poles of $\chi(\omega,k)$ shown in \textsf{a} and \textsf{c}. The dashed curve  in \textsf{b} shows a fit to $\xi = \xi_0|\epsilon|^{-\nu}$  resulting in $\xi_0 = 0.28 \pm 0.02$ and $\nu=0.500\pm0.002$. The dashed curve in \textsf{d} shows a fit to $\tau = \tau_0 |\epsilon|^{-z\nu}$, resulting in $\tau_0 = 2.02\pm 0.01$. Furthermore, we can analytically determine $z=2$. Within numerical accuracy, we find identical values for $z$ and $\nu$ below the critical point, with slightly different $\xi_0 = 0.38\pm 0.01$ and $\tau_0 = 0.89$. We have the approximate relation $\xi_0^{<} = \sqrt{2}\xi_0^> $ between $\xi_0$ below and above the transition. This is the opposite of the usual Landau-Ginzburg mean-field relation $\xi_0^{<} = 1/\sqrt{2}\xi_0^> $. Thus, while the critical exponents have the mean-field values, equilibrium correlations lengths clearly show that the theory we are dealing with is outside of the the Landau-Ginzburg paradigm.\label{fig.UniversalityClass}}
\end{center}
\end{figure}

Adapting the analysis of \cite{Maeda:2009wv} to the present situation, the correlation function of the order parameter field $\cO$ near, but slightly above, $T_{\rm c}$ takes the momentum-space form
\be\label{eq.DynamicResponse}
\langle \cO(\omega,k)\,\cO^\dagger(-\omega,-k)\rangle :=\chi(\omega,k) = \frac{Z(\omega,k)}{ic \omega + k^2 + \frac{1}{\xi^2}}\,,
\ee
where the last expression holds for small $\omega,k$ and we introduced the parameters $c$ and $1/\xi^2$.
We have defined the dynamical susceptibility of the order parameter $\chi(\omega,k)$ in the first equality. One obtains the correlation function from the susceptibility $\chi(\omega,k)$ by a Fourier transform
\be
G\left({t-t'},x-x'\right) = \int \frac{d\omega dk}{(2\pi)^2} e^{i \omega (t-t')-ik(x-x')}\chi(\omega,k)\,.
\ee 
We can find the relaxation time by looking at the Fourier transform of the zero-momentum response, $\chi(\omega,k=0)$, which at late times takes the form
\be\label{eq.DynamicLongTime}
G({t-t'}) \propto e^{-\frac{t-t'}{c \xi^2}}\,,\qquad t\gg t'\,,\qquad \Rightarrow\tau = c\xi^2\,.
\ee
The equal time correlation function, following from the Fourier transform of the static correlation function $\chi(\omega=0,k)$, falls off like
\be
G(t=t',x-x') \propto e^{-\frac{x-x'}{\xi}}\,,\qquad x\gg x' \,,\qquad \Rightarrow z=2\,.
\ee
From the fact that the relaxation time is proportional to the square of $\xi$ it follows that the dynamical critical exponent $z=2$. This leaves us to determine the correlation length critical exponent $\nu$. For this we must study how the correlation length $\xi$ diverges near the critical point, $\xi \sim \xi_0 |\epsilon|^{-\nu}$. Using the above relations \reef{eq.DynamicResponse} \& \reef{eq.DynamicLongTime}, we deduce that $\tau \sim \tau_0 |\epsilon|^{-2\nu}$ with $\tau_0 = c \xi_0^2$. In holography the poles of two-point functions of boundary operators coincide with the quasinormal modes of the bulk fields dual to the operators appearing in the correlation function \cite{Kovtun:2005ev}. As we saw above, in order to extract $\xi$ and $\tau$ it is sufficient to study the static susceptibility $\chi(\omega=0,k)$ and the dynamical susceptibility $\chi(\omega,k=0)$ separately. As shown in Fig. \ref{fig.UniversalityClass}, we find that the spectrum of quasinormal modes of $\chi(\omega,k=0)$  contains poles only in the lower half complex plane (as demanded by stability), while $\chi(\omega=0,k)$ has two series of poles along the positive and negative imaginary axis. This structure is also apparent from the correlation function \reef{eq.DynamicResponse}. In each case the relevant poles are the ones closest to the real axis, governing the exponential decay at long time scales (defining the relaxation time $\tau$), and the falloff of correlations at large distances (defining the correlation length $\xi$), respectively. We find from the quasinormal mode analysis that $1/\xi^2 \sim |\epsilon|$, which implies $\nu = 1/2$, consistent with the results of \cite{Maeda:2009wv}. By studying the motion of the leading poles as $T\rightarrow T_{\rm c}$ as discussed in Methods,  we can also deduce that $\tau_0=2.02\pm 0.01$ and $\xi_0=0.28\pm 0.02$ if the critical point is approached from above and $\tau_0=0.89 \pm 0.01$ and $\xi_0=0.39\pm 0.01$ if it is approached from below. For more details, we refer the reader to Fig.~\ref{fig.UniversalityClass}.

\pagebreak

\clearpage

\subsection{Far from equilibrium dynamics}

\begin{figure}[h!]
\begin{center}
\includegraphics[width=0.95\textwidth]{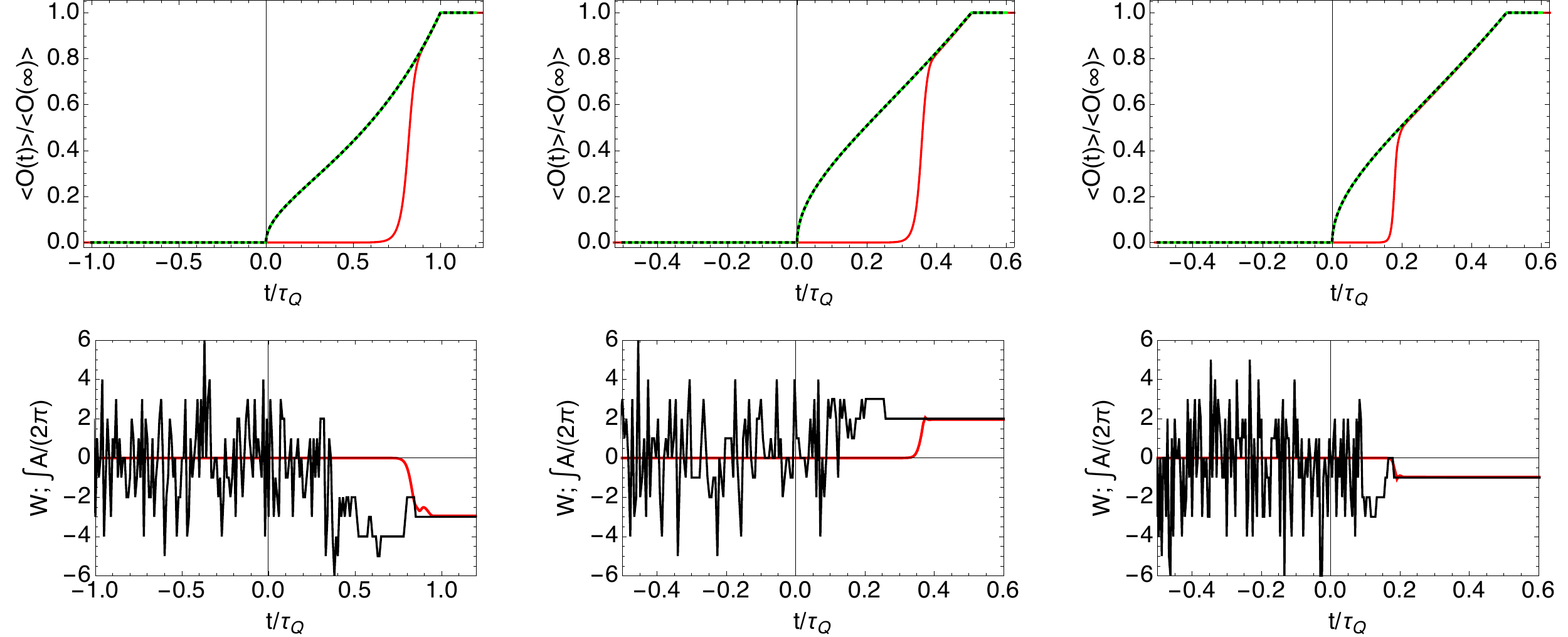}
\begin{picture}(0.1,0.25)(0,0)
\put(-155,35){\makebox(0,0){\textsf{a}}}
\put(-100,35){\makebox(0,0){\textsf{b}}}
\put(-47,35){\makebox(0,0){\textsf{c}}}
\put(-155,2){\makebox(0,0){\textsf{d}}}
\put(-100,2){\makebox(0,0){\textsf{e}}}
\put(-47,2){\makebox(0,0){\textsf{f}}}
\end{picture}
\caption{{\bf Time-evolution of the order parameter and the winding number.} Representative samples of individual runs for three different quench rates. First column (panels \textsf{a} \& \textsf{d}): $\tau_{\rm Q} = 10$; second column (panels \textsf{b} \& \textsf{e}), $\tau_{\rm Q} = 35$; third column (panels \textsf{c} \& \textsf{f}): $\tau_{\rm Q}=120$. In the first row (\textsf{a} - \textsf{c}) we show the time development of the order parameter (red curve) compared to the instantaneous adiabatic value (green-black dashed). In the second row (\textsf{d} - \textsf{f}) we show the time evolution of the winding number (black) compared to the line integral of the gauge field $\frac{1}{2\pi}\oint A$. The two converge in equilibrium, as dictated by gauge invariance: $\oint d\Theta = \oint A$. The winding number is always integer quantized, the non-vertical parts of the curve are an artifact of joining up the quantized values at finite sampling intervals. The winding number is physically well defined only once a condensate has developed. \label{fig.realTime}}
\end{center}\end{figure}
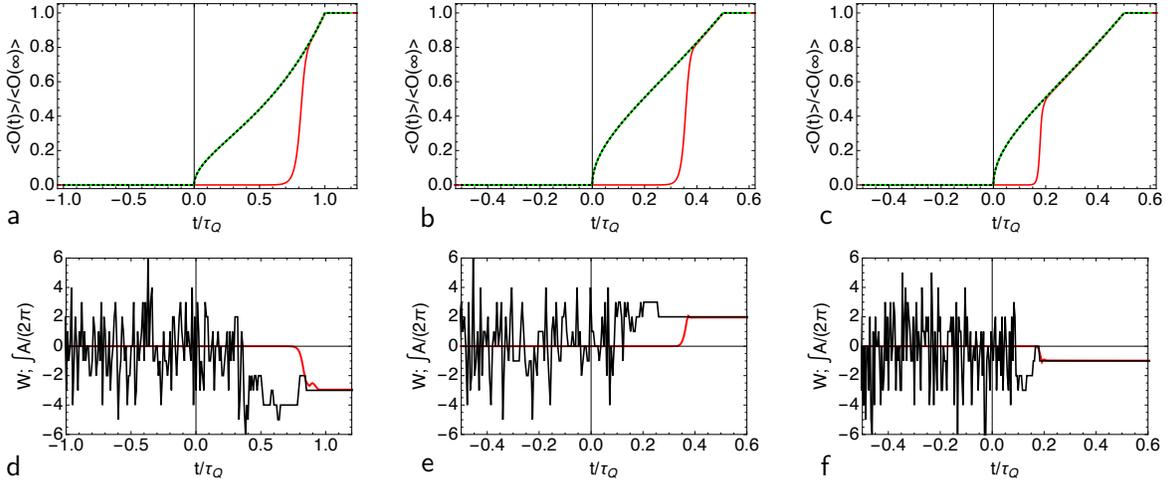

We simulate cooling of the superconducting ring by numerically evolving the bulk equations (Eqs. \reef{eq.eom} in Methods) in the black-hole background with changing temperature \reef{eq.Thawk}. We implement a piece-wise linear protocol starting at an initial temperature of $T_i$ with corresponding $\epsilon_i$,  and cooling the system at finite rate according to $\epsilon(t) = -t/\tau_{\rm Q}$ through the critical point, until the temperature $T_f$ with corresponding $\epsilon_f$ is reached. We implement the temperature ramp by changing the dimensionless ratio $T C = C/2\pi z_{\rm h}$, while holding the density $\rho C$ fixed. For a precise definition of these quantities the reader may consult the Methods section. In order to allow the system to break the symmetry dynamically, one needs to add fluctuations to the classical Einstein equations. To achieve this, we introduce noise into the evolution in a manner consistent with the fluctuation-dissipation theorem for the horizon temperature $T_H$. Thus we sample the boundary values, $\psi(z,t,\phi)\bigr|_{z=z_{\rm UV}}$, of the scalar field (its real and imaginary parts) from a Wiener process, which ensures that their average values vanish but its dynamics gives rise to a non-vanishing correlation. We treat the amplitude of the noise as a phenomenological parameter, but it would be enlightening to determine its precise form in the future, for example by deriving the relevant fluctuation-dissipation relation from bulk quantum effects (for work in this direction, see \cite{deBoer:2008gu,Son:2009vu,Sonner:2012if}). Before the linearly decreasing temperature ramp we allow the system to thermalize for a fixed time in order to allow the noise introduced at the boundary to get distributed over all scales in the bulk. 
In order to solve the nonlinear partial differential equations determining the evolution of the bulk fields, we use a Chebyshev grid in the holographic (`radial') direction $z$ and a Fourier decomposition in the periodic direction $\phi$. We use a fixed step to evolve in time. Each quench is started at some $\epsilon > \hat \epsilon$ for a given $\tau_{\rm Q}$ and is stopped at a time, $t_{\rm stop}$ when the condensate, averaged over the ring, has reached a fixed fraction of the equilibrium value at $\epsilon(t_{\rm stop})$. The time elapsed between the time of crossing the equilibrium phase transition point, and the stopping time $t_{\rm stop}$ defines the lag time $t_{\rm L}$. At this point the winding number $W$ is recorded together with the lag time $t_{\rm L}$. For the simulations of this work, a fraction of $0.9$ was employed. We found that the winding number became frozen, i.e. time independent, before that time. Both of the recorded quantities are predicted in the KZM scenario to follow the universal scaling relations \reef{eq:hat_t2} and \reef{eq.Wscale} once averaged over noise realizations. The winding number $W$ is extracted from the discretized $\phi$ direction (denote the grid points by $\{ \phi_i\}_{i=1}^{N_x}$) via the sum
 \be\label{eq.noiseCorrelation}
 W = \frac{1}{2\pi} \sum_{i=1}^{N_x} {\rm Arg}\left(\cO_{i+1} \cO_i^\dagger \right)\,.
 \ee
 As in the case of the continuum definition above, the sum as a whole is gauge invariant under single-valued gauge transformations.  We find that it stabilizes soon after the ordered phase is formed (see Fig. \ref{fig.realTime}). Winding number becomes a good observable by the time we stop the evolution to extract both $t_{\rm L}$ and $W$, when it is frozen in at integer values. From this point on it is no longer susceptible to noise or late-time equilibration dynamics of the ordered phase, as can be seen in Fig. \ref{fig.realTime}. 
 
 An example condensation process for $\tau_{\rm Q}=12$ is shown in Fig. \ref{fig.Summary}, where boundary and bulk physics leading to a $W=3$ configuration is illustrated.

We find very good agreement between the full simulations and KZM predictions based on the equilibrium critical exponents deduced from our independent quasinormal mode analysis. From $\nu = 1/2$ and $z=2$ it follows that $\langle |W|\rangle \propto \tau_{\rm Q} ^{-1/8}$ and $t_{\rm L} \propto\tau_{\rm Q}^{1/2}$, while the simulation results in
 \be
 \langle |W|\rangle \propto \tau_{\rm Q}^{-0.13 \pm 0.02}\,,\quad t_{\textrm{L}} \propto \tau_{\rm Q}^{0.46 \pm 0.005}\,.
 \ee
 The quoted uncertainties give a single standard deviation from the fitted value.
The discrepancy in the value of $t_{\rm L}$ is likely a result of the fact that the lag time does not vanish at very small quench times, but rather saturates to a finite value, so that the simple scaling form is no longer a very good fit for the rapid quenches at the fast end of our window of $\tau_{\rm Q}$.
 In summary, we find good agreement with universal KZM values for the scaling of $t_{\rm L}$ as well as the dispersion of winding number $\sigma \left( W \right)$. Matching the prefactors in \reef{eq:hat_xi} with equilibrium predictions usually results in more significant quantitative departures. In the present case we obtain
 \be
\xi_0^{\rm sim} = 1.16^{\pm 0.21}\,,\qquad \tau_0^{\rm sim} = 3.92^{\pm 0.12}\,,
 \ee
 deviating by a factor of $\sim 3$ from the equilibrium values ($\xi_0 = 0.4$ and $\tau_0 = 2.02$) extracted from correlation functions. From past experience it is to be expected that the KZM values over-estimate the number of defects.  KZM predictions of $\sigma(W)$  for $\tau_{\rm Q} = 10, 100, 1000$  are $2.44, 1.83, 1.37$ (rounded to two significant digits), compared to $1.44, 1.31,  1.00$ (again, rounded) from the full simulation, so KZM overestimates the density of defects. Evidently the numbers in our simulation are in rather good agreement, compared to past simulations where mismatches by factors of ${\cal O}(10)$ were not uncommon (see e.g. \cite{das2012winding}). 
Provided that the numerics is good enough, the degree of agreement depends on the microscopic dynamics.
 Furthermore our value for $\tau_0$ is in line with the mismatch encountered in previous work \cite{das2012winding}. 
\pagebreak
\clearpage
 \begin{figure}[h!]
\begin{center}
\hskip1em\textsf{a}\includegraphics[width=0.45\textwidth]{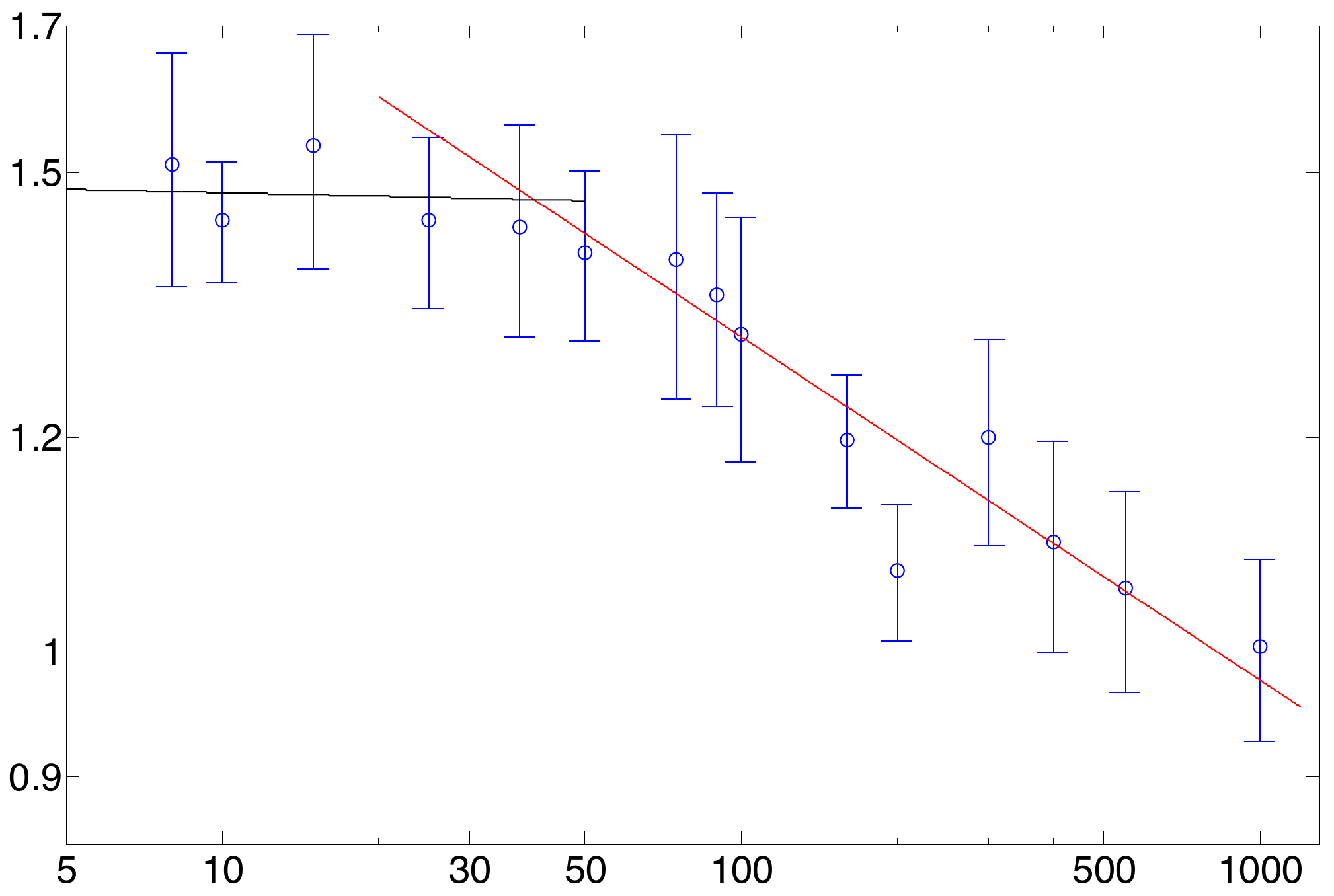}\begin{picture}(0.1,0.25)(0,0)
\put(-30,-2){\makebox(0,0){$\tau_{\rm Q}$}}
\put(-75,32){\makebox(0,0){\begin{sideways}$\sigma(W)$\end{sideways}}}
\end{picture}\vskip3em
\textsf{b}\includegraphics[width=0.45\textwidth]{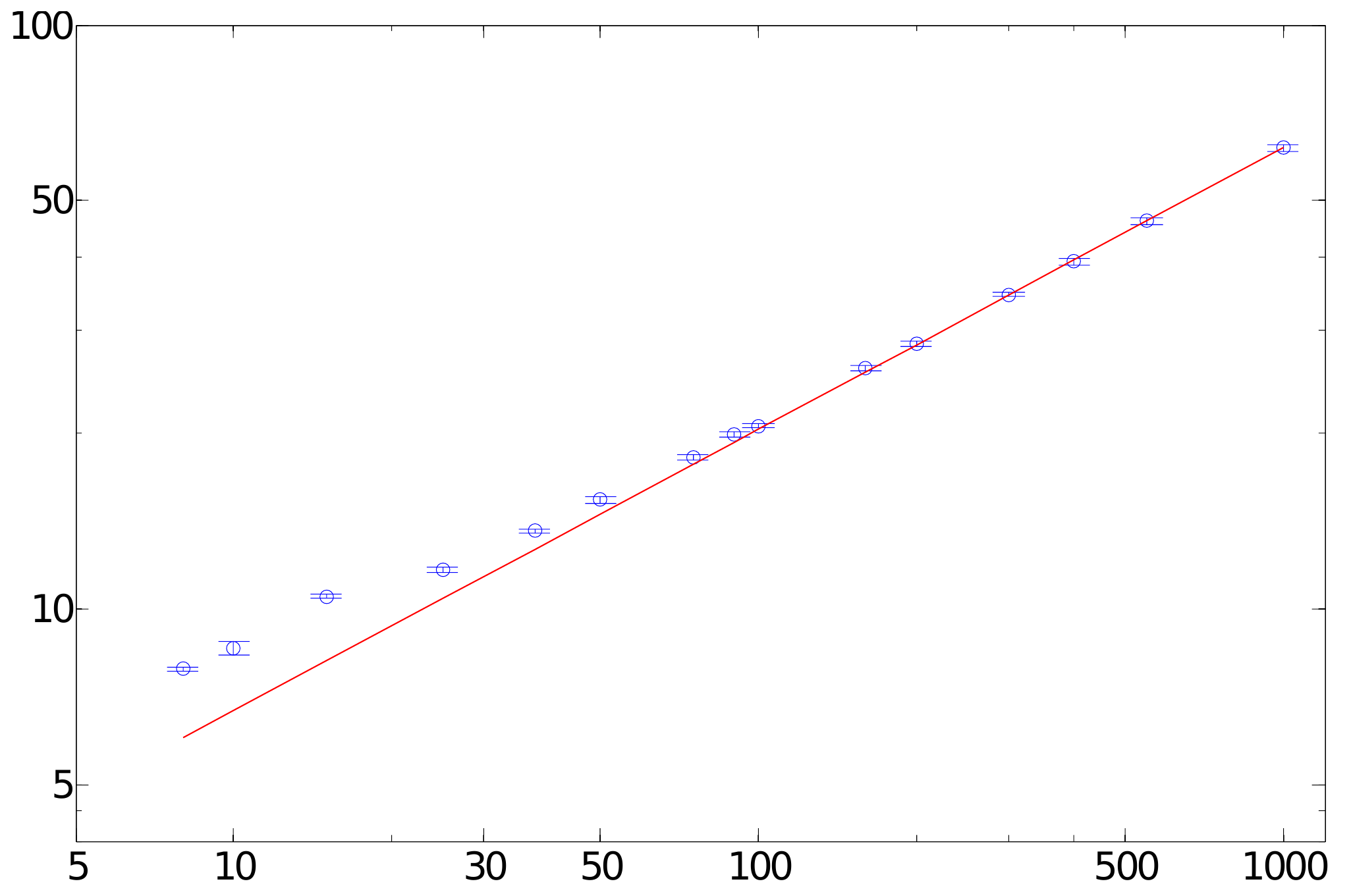}\begin{picture}(0.1,0.25)(0,0)
\put(-35,-2){\makebox(0,0){$\tau_{\rm Q}$}}
\put(-75,27){\makebox(0,0){\begin{sideways}$t_{\rm L}$\end{sideways}}}
\end{picture}\hskip2em
\textsf{c}\includegraphics[width=0.47\textwidth]{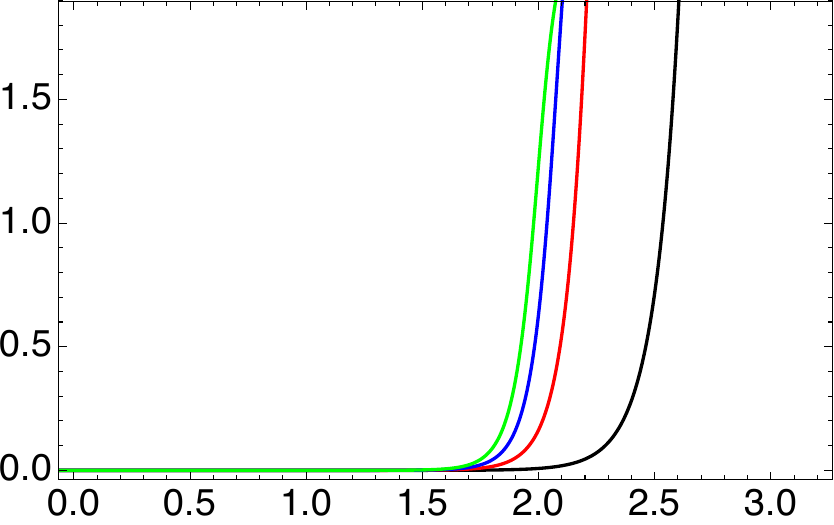}\begin{picture}(0.1,0.25)(0,0)
\put(-34,-2){\makebox(0,0){$t/\hat t$}}
\put(-80,27){\makebox(0,0){\begin{sideways}{\small $\langle {\cal O}(t)\rangle$}\end{sideways}}}
\end{picture}\
 \caption{{\bf Universal scaling laws for the lag time and the dispersion of winding number.} Panel \textsf{a}: Best fit to winding number scaling gives excellent agreement with values predicted by KZM. Shown is a fit to $\sigma(W)$, which gives $\sigma(W) = a \tau_Q^b$, with $a = 2.34 \pm 0.25$ and $b = -0.126\pm 0.02$, which should be compared to the prediction $\gamma_0 \tau_{\rm Q}^{-1/8}$, where equilibrium estimates for $\gamma_0$ give the value $3.99$. The error bars in panel \textsf{a} were computed as the standard deviation of the mean for each $\tau_{\rm Q}$ in our dataset for winding numbers. The error bars in panel \textsf{b} give a single standard deviation from the mean for each $\tau_{\rm Q}$ in the dataset. We see that the dispersion in winding number saturates for rapid quenches. These deviations are accompanied by a deviation from the naive scaling prediction of $\hat t$ at the onset of saturation for $\sigma(W)$ (see also panel \textsf{c} and its description below). Panel \textsf{b}: Best fit to scaling of the lag time $t_{\rm L}$. Inset shows the condensate as a function of $t/\hat t$ for the three cases of Fig. \ref{fig.realTime}.  Panel \textsf{c}: Order parameter ${\cal O}(t)$ averaged over the ring as a function of time, given in units of $\hat t$. This last quantity is computed from the near-equilibrium critical behavior of our system. We show four different values of the quench time, $\tau_Q = 10$ in black, $\tau_{\rm Q}=35$ in red, $\tau_{\rm Q}=120$ in blue, and finally $\tau_{\rm Q} = 300$ in green. The scaling collapse for lag time becomes more and more accurate for slower quenches, in accordance with the results in panels \textsf{a} \& \textsf{b}, showing more significant deviations for fasts quenches. \label{fig.wind}}
\end{center}\end{figure}
\clearpage
\pagebreak
\section{Discussion}
The work reported here constitutes the first demonstration of universal KZ scaling of defects spontaneously created in the far-from equilibrium dynamics of a strongly-correlated field theory without quasiparticles. Holography allowed us to map the time-dependence of this system to the evolution of a set of nonlinear partial differential equations together with stochastically sampled boundary conditions. Our setup is comparable to a holographic version of the Gross-Pitaevskii equation, applicable for systems without quasiparticles. 

We introduced stochasticity by sampling boundary conditions from a Wiener process characterized by a phenomenological parameter $\alpha$. This is natural, because it constitutes only a minimal extension of the usual holographic dictionary, which tells us how to map bulk to boundary quantities, but has the limitation that there is a free parameter. While we found that the universal results concerning the scaling law as a function of the cooling rate are very robust to changes in $\alpha$, and the parameters $\tau_0^{\rm sim}$ and $\xi_0^{\rm sim}$ of our study show only weak dependence, it would nevertheless be desirable to derive it from first principles. This means deriving the fluctuation-dissipation relation associated with the thermal Hawking radiation of the bulk black hole, which will force us to take into account quantum effects in the bulk gravity model along the lines of \cite{deBoer:2008gu,Son:2009vu,Sonner:2012if}.

Our scaling results are consistent with exponents obtained in the mean-field approximation of the boundary theory, even though the theory is not described by a simple Landau-Ginzburg (LG) model, as evidenced e.g. by a violation of the standard LG relation between correlation length above and below the transition (see Fig. \ref{fig.UniversalityClass} and its caption). The mean-field like scaling results are a consequence of working in the classical gravity limit, which, however, does manage to capture some features beyond mean-field, notably the ratio of correlation lengths above and beyond the phase transition. In holography, this quantity takes the form of a ratio of bulk integrals \cite{Maeda:2008ir}, and thus receives contributions from all scales, including the horizon, the source of dissipation in our holographic representation. The fact that dissipative dynamics are naturally incorporated into the theory from first principles makes holography a powerful tool to investigate strongly-coupled superfluid and superconductor dynamics, as previously stressed in \cite{Adams:2012pj}, where the holographic origin of dissipation was crucial in establishing a turbulent direct cascade. In more conventional approaches, such as the stochastic Gross-Pitaevskii or Landau-Ginzburg approaches, dissipation has to be added by hand. A tempting next goal would be to investigate quench dynamics in theories that do not result in mean-field scalings.  A much more ambitious (but perhaps not completely unrealistic \cite{McGreevy:2010}) goal would be to study strongly coupled theories directly relevant to condensed matter physics.

An open question left unanswered by the present investigation is the origin of the saturation we observed. There are several plausible explanations, and we hope to return to a detailed investigation of the reasons for the saturation in the model we have presented here in forthcoming work.

Another interesting future project would be to study the effect of the noise in more detail. In addition to the first principles derivation of the properties of the Wiener process we have already noted, two distinct phenomenological tacks can be considered. We have already begun to investigate the effect of the amplitude of the Wiener process (our parameter $\alpha$) applied at the boundary. The preliminary conclusion is that -- at least within the range we have explored -- the quantities we have followed are insensitive to suitably small $\alpha$, with the dependence likely to be sublogarithmic. More detailed characterization of this dependence would be desirable. Furthermore, one can consider noise applied throughout the AdS interior (rather than only on the boundary). This raises the possibility of seeding topological defects in the interior, and may pose the question of the relation between them and the behavior of the field on the boundary.

\noindent
{\bf Acknowledgements:}\\
\noindent
We would like to thank Allan Adams, Tarek Anous, Chris Herzog, Nabil Iqbal, Kristan Jensen, Arttu Rajantie, Homer Reid and Toby Wiseman for very helpful discussions. The numerical computations in this paper were performed on the MIT LNS openstack cluster, and we thank Jan Balewski and Paul Acosta for their kind assistance. We also thank Allan Adams for providing us with further computational resources.
This research is supported by the U.S Department of Energy through the LANL/LDRD Program and a  LANL J. Robert Oppenheimer fellowship (AdC). This work was also supported in part by the U.S. Department of Energy (DOE) under cooperative research agreement Contract Number DE-FG02-05ER41360 (JS).

\section{Methods}\label{sec.holoRenorm}
\subsection{Bulk action and Details of the Holographic Mapping}
The full action for the AdS$_3$ holographic superconductor reads
\be\label{eq.bulkAction}
S = \frac{1}{16\pi G_N}\int d^3x\sqrt{-g} \left( R + \frac{2}{\ell^2}\right) - \frac{1}{q^2}\int \sqrt{-g} d^3x\left(\frac{1}{4}F^2 + |D\psi|^2  + m^2 |\psi|^2 \right)\,,
\ee
where $D_\mu \psi = \nabla_\mu\psi - iA_\mu\psi$ and we choose the case $m=0$. We denote bulk indices running over $z,t,x$ by $x^\mu$ and boundary indices, running over $t,x$, with $x^i$. Taking $q\rightarrow\infty$ has the result of decoupling the gravity part from the matter part \cite{Hartnoll:2008vx}, tantamount to the `probe limit' discussed in this article. One fixes a solution of the gravity equations, to be discussed shortly, and treats the dynamics of the matter fields separately. Hence the effective bulk action from which the equations of motion follow is simply Maxwell-scalar theory in a non-trivial background, that is, the equations of motion are
\be\label{eq.eom}
D^2 \psi = 0\,,\qquad \nabla_\mu F^{\mu\nu}+ J^\nu=0\,,
\ee
for the current $J_\mu = i(\psi (D_\mu\psi)^* - \psi^* D_\mu \psi )$. Covariant derivatives are taken with respect to a fixed background metric, which we take to be the metric of the BTZ black hole, written in ingoing Eddington-Finkelstein coordinates
\be\label{eq.BTZmetric}
ds^2 = \frac{\ell^2}{z^2}\left( -f(z) dt^2 - 2 dt dz + dx^2 \right)\,.
\ee
The coordinate $x$ is identified $x\sim x + C$ for the dynamical simulations, where for numerical convenience a rescaled coordinate $\phi =2\pi x/C $ is used. The Eddington-Finkelstein time coordinate $t$ reduces to boundary time at $z=0$, warranting the use of the same symbol for both. The coordinate $x$ is kept unidentified for the calculations of the quasinormal modes in section D
.
The length $\ell$ sets the AdS curvature scale; its role is to fix the number of degrees of freedom `$N$' of the dual theory, which must be large for classical gravity to apply. At leading order in large $N$ this quantity scales out of the equations we solve. The metric function takes the fixed form $f = 1-z^2/z_{\rm h}^2$.
 For the scalar field one finds the asymptotic expansion
\be
\psi(t,z,x) = \psi_{(0)}(t,x) + \psi_{(2)}(t,x) z^2 + \cdots
\ee
The interpretation of $\psi_{(0)}(t,x)$ is that it sources the symmetry-breaking operator $\cO (t,x) $, while $\psi_{(2)}(t,x)$ gives its expectation value
\be \psi_{(2)} =\frac{1}{2\ell} \langle \cO\rangle\, .\ee
 Therefore the requirement that the symmetry be broken spontaneously means that we must set the source $\psi_{(0)}(t,x)$ to zero for all time. This translates into a homogenous Dirichlet boundary condition on the field $\psi(t,z,x)$ in the UV,
 \be\label{eq.Dirichletpsi}
 \psi(t,z,x)\Bigr|_{z=z_{\rm UV}}=0\,.
 \ee
  The vector field in AdS$_3$ is more subtle \cite{Marolf:2006nd}. Its asymptotic behavior is given by
 \be\label{eq.GaugeUV}
 A_\mu(t,z,x) = j_\mu(t,x) \log(z/\Lambda) + a_\mu(t,x) + \cdots\,,
 \ee
 where the vector $j_\mu$ is an external current in the boundary theory, not to be confused with the bulk current $J_\mu(t,z,x)$. We introduced the scale $\Lambda$ to make the argument of the logarithm dimensionless.  We shall see that our chosen boundary condition on $A_\mu$ is independent of this scale. Note that in the normal phase the solution is $j_\mu \log(z/z_{\rm h})$ and $\Lambda = z_{\rm h}$ is enforced by regularity at the horizon. It is convenient to work in axial gauge so that $A_z=0$ and thus the current $j_\mu(t,x)$ has components only in the field theory directions, $j_i(t,x)$. Note that this is not the same as choosing axial gauge in the Schwarzschild like coordinate system \reef{eq.SchwarzschildForm}, in which the bulk metric is diagonal. In the latter choice of coordinates, the equations of motion imply the equation
 \be\label{eq.CurrentWardIdentity}
 \partial_i j^i_S - i \ell \left(\psi_{(0)}\langle \cO \rangle^* - \psi_{(0)}^* \langle \cO \rangle \right)=0\,.
 \ee
Thus the current is conserved in the absence of a source for the operator $\cO(t,x)$ (We have denoted the current evaluated in the Schwarzschild like coordinates as $j_S^i$). From the point of view of the boundary field theory this is simply the Ward identity for the one-point function of the current. Operationally, the conservation condition follows from the $z-$component of the Maxwell equations, which gives rise to a constraint.
 Turning to the gauge field, the boundary condition is
  \be\label{eq.GaugeNeumann}
 \Pi_A^\mu\Bigr|_{z=z_{\rm UV}} + j^\mu =0\quad \Rightarrow\quad \frac{z}{\ell}\partial_z A_\mu\Bigr|_{z=z_{\rm UV}}=j_\mu\,,
 \ee
 where $\Pi^\mu_A$ is the momentum conjugate to $A_\mu$ with respect to $z$ slicing
 \be
 \Pi^\mu_A = \lim_{z\rightarrow z_{\rm UV}} \sqrt{-g}F^{\mu z}\,.
 \ee
 This is a Neumann boundary condition. Thus we are free to fix $j_i(t,x)$ subject to the conservation condition, and leave $a_i(t,x)$ free to fluctuate. As stated above, the scale $\Lambda$ drops out from the boundary condition \reef{eq.GaugeNeumann}. This choice corresponds to a dual vector operator of dimension $\Delta=1$, the right dimension for a dynamical gauge field in the boundary theory. Indeed the residual gauge transformation preserving axial gauge ($\lambda_{\rm res}(t,z,x) = \lambda(t,x)$) acts on this as a standard field-theory gauge transformation
\be
a_i(t,x) \rightarrow a_i(t,x) + \partial_i \lambda(t,x)\,,
\ee
so that $a_i(t,x)$ is a bona-fide fluctuating gauge field. We fix a constant background charge density, so that the current has the only non-vanishing component, $j^t = \rho$.
\subsection{Details on Numerics}
The simulations in this article were performed on a pseudo-spectral spatial grid comprised of 21 Chebyshev points in the radial direction and 111 plane waves in the angular direction of the boundary. 

For each run we start the system at the initial time slice in the normal phase, defined by setting to zero the field $\psi$ and giving the gauge field a non-trivial time component $j^t=\rho$. We then evolve forward in time, updating the noise according to the rule \reef{eq.noiseCorrelation}, implemented as a discrete Wiener process. We average over ${\cal O}(10^2)$ noise realizations for each value of $\tau_{\rm Q}$ to compute  $\sigma(W)$. Computer codes used in the simulations of this work are available upon request.

\subsection{Evolution Scheme}
For the simulations reported on in this article we employed a characteristic evolution scheme for the Maxwell and scalar fields. After gauge fixing $A_z=0$, denoting $A_t(t,z,x) = T(t,z,x)$, $A_x(t,z,x) = X(t,a,x)$ and writing $\psi(t,z,x) = a(t,z,x) + i b(t,z,x)$ we have four evolution equations and one constraint equation. In the numerical evolution it proves convenient to work with rescaled fields, rather than the `bare' ones appearing in the action \reef{eq.bulkAction}
. We furthermore subtract the leading log terms. Thus defining $\hat T = z \left(T + \rho \log z\right)$ and $\hat X = z X$ we find the equations
 \bea\label{eq.evolve}
 \left[\partial_{z} - \frac{1}{2z}\right]\Phi^i_t &=& S^i[a,b,T,X]\,, \nn
  \left[\partial_{z} - \frac{1}{z}\right]\hat{T}_t&=& S^T[a,b,T,X]\,,
 \eea
 where the fields $\Phi^i$ are all fields other than $\hat T$. The sources $S^i$ and $S^T$ depend non-linearly on the fields as well as their spatial derivatives. In addition, the radial ($z$) component of the Maxwell equation gives the constraint equation
 \be
\left(1 - z \partial_z + z^2 \partial^2_z  \right)\hat T = 2\ell^2z \left( b\partial_z a - a \partial_z b \right) + z^2 \partial^2_{zx}\hat X- z \partial_x \hat X\,.
 \ee
 After the rescaling, the boundary conditions on $\hat T$ and $\hat X$ are now
 \be
 \hat T(t,z,x)\Bigr|_{z=0} = \hat X(t,z,x)\Bigr|_{z=0}=0\,.
 \ee
 Note that Eqs. \reef{eq.evolve} are ${\it linear}$ equations for the time derivatives suggesting the following evolution scheme. Assume that at time $t=t_n$ the values of the fields $a,b,\hat X$ are known. We can now use the constraint equation to solve for $\hat T(t_n,z,x)$ and then solve the linear equations for $x^i_t$ to obtain the time derivatives of $a,b,\hat X$. We can then use our favorite time evolution scheme to obtain the values of $a,b,\hat X$ on the next time slice $t_{n+1}$. We have used explicit RK4 integration as well as simple forward Euler with good results. For $111$ Fourier modes in the spatial boundary direction, the step size was taken to be $0.003/N_z$, where $N_z$ is the number of grid points in the radial direction. The scaling analysis of Figs. 5
 ~was obtained on a grid of $21$ points in the radial direction and $111$ Fourier modes in the annular direction choosing a characteristic size of the ring $C = 50\ell$. It would be desirable to repeat the analysis for a larger ring with higher spatial resolution, which is likely to require more significant computing power, especially if higher statistics on noise realizations is desired.
 
 The linear equations for $\partial_t\Phi^i_n$ on a given time slice can then be solved efficiently and in parallel for each spatial grid point $\phi_j$. For the above-mentioned spatial grid we choose a fixed time step of $\Delta t = 0.003/N_z$. In order to achieve stable long-time evolution we filter out high-frequency modes in the $\phi$ direction using the Orszag $2/3$ rule every five time steps. This can be achieved very efficiently using FFT and iFFT. Due to the presence of logarithmic terms in the asymptotics of the fields one does not expect spectral accuracy, and this is borne out in preliminary convergence tests. We  found that our evolution scheme is stable even in the presence of stochastic boundary conditions simulating thermal noise and gives accurate results (for example by comparing the dynamical solutions at late time to the equilibrium results obtained in the standard way from solving ODEs). We now turn to a detailed description of our implementation of stochastic boundary conditions.


\subsection{Holographic Renormalization}
In axial gauge, and setting $\psi_{(0)}=0$, the equations of motion \reef{eq.eom}
 have solutions with asymptotic behavior
\bea
A_i(z;t,x) &=& j^{(0)}_i(t,x) \log\left( z \right) + a^{(0)}_i(t,x) +\cdots \nn
 \psi(z;,t,x) &=&  z^2 \psi^{(2)}(t,x) + \cdots
\eea
We regularize the on-shell action by introducing a finite cutoff $z = \epsilon$, so that we now have
\begin{widetext}
\bea
S &=& -\int_{z = \epsilon}d^2 x \sqrt{-\gamma} \hat n_\mu\left[ \tfrac{1}{2}g^{\mu\nu} \left( (D_\nu \psi) \psi^\dagger + \psi D_\nu \psi^\dagger \right) + F^{\mu\nu}A_\nu\right] + {\rm E.O.M.} + \cdots\,,\nn
&=&-\frac{1}{2\ell}\int_{z=\epsilon} dt dx \left[ j^{(0)}_i(t,x)^2 \log \epsilon + j^{(0)}_i(t,x) a^{(0)}_i(t,x) \right] + \cdots
\eea
where $\hat n$ is the outward pointing unit normal to the boundary and $\gamma = {\rm det}(\gamma_{ij})$ the determinant of the induced metric. The omitted terms depend only on the boundary values $j_i, a_i, \psi^{(2)}$ and vanish in the limit $\epsilon \rightarrow 0$. The divergent terms can be cancelled by adding a counterterm
\be
S_{\rm CT} = \frac{1}{2 \ell} \int d^2 x \sqrt{-\gamma}F_{zi}F^{zi}\log\epsilon\,.
\ee
As shown in \cite{Jensen:2010em} this counterterm becomes a manifestly local boundary quantity when written in terms of a dual scalar field $S$, obtained from the two-form gauge field $F$ via $F = \star d S$.
\end{widetext}

 \subsection{Noise}
 Under classical evolution, in the case at hand corresponding to the classical large $N$ limit of the dual field theory, fluctuations are suppressed and the symmetry of the order parameter cannot be dynamically broken. Said differently, even though below $T_c$ the phase with $\langle \cO \rangle = 0$ is unstable, there is nothing in the classical evolution equations to push the order parameter off its precarious perch on top of the potential. Taking our lead from the literature on BEC dynamics using the stochastic Gross-Pitaevksii equation \cite{sabbatini2011phase} we add fluctuations by sampling our boundary conditions from a thermal noise distribution, that is from a Wiener process. Thus the boundary conditions for $a(t,z,x)$ and $b(t,z,x)$ should be modified. Instead of imposing the strict Dirichlet boundary condition \reef{eq.Dirichletpsi}
 , we only impose this condition on average, i.e. we impose $\langle \langle a_{(0)}(t,x)\rangle \rangle = \langle \langle b_{(0)}(t,x)\rangle \rangle  =0$, with
 \bea\label{eq.noisyBoundary}
 \langle \langle a_{(0)}(t,x) a_{(0)}(t',x')\rangle \rangle &=& \alpha \frac{T}{T_c} \delta(t-t') \delta(x-x')\,,\nn
  \langle \langle b_{(0)}(t,x) b_{(0)}(t',x')\rangle \rangle &=& \alpha \frac{T}{T_c}\delta(t-t') \delta(x-x')\,,
 \eea
 where $\langle \langle \cdot \rangle \rangle$ denotes noise average and $\psi_{(0)} = a_{(0)} + i b_{(0)}$. Usually one determines $\alpha$ from a fluctuation-dissipation relation as $\alpha = 2\eta T_c$, where $\eta$  is a damping parameter. In this work we treat $\alpha$ as a phenomenological parameter. We found that varying $\alpha$, even by several orders of magnitude has little influence on the scaling results, but we do find a weak dependence of the absolute magnitude of $t_{\rm L}$ on $\alpha$. Clearly it is desirable in future to determine $\alpha$ from a first-principles holographic calculation.

 In practice the noisy boundary condition is realized by sampling each spatial boundary point from an independent normal distribution of zero mean and unit variance $N(0,1)$. That is, we set
 \bea\label{eq.noisyBoundary}
 a_{(0)}(t_n,x_i) &=& \alpha\frac{T}{T_c} \sqrt{\Delta t} N(0,1)\,,\nn
 b_{(0)}(t_n,x_i) &=& \alpha \frac{T}{T_c} \sqrt{\Delta t} N(0,1)\,,
 \eea
 for each boundary point $x_i$. Note that one cannot choose both the boundary value of $j_i$ and $\psi_{(0)}$ independently, since they are constrained by the current Ward identity (26) 
 .
This equation is consistent only for $\psi_{(0)}=0$, so in what sense can one choose the noisy boundary condition above? This can be seen by considering the noise as a small perturbation $j_i + \delta j_i$ and $\psi_{(0)} + \delta \psi_{(0)}$ satisfying
\be\label{eq.noiseCurrent}
\partial_i \delta j^i -i \ell \left( \delta\psi_{(0)}  \langle\cO \rangle^* - \delta\psi_{(0)}^* \langle \cO \rangle\right) =0\,.
\ee
 By solving the $z$ component of the Maxwell equation near the boundary we automatically impose this constraint at each time step, that is the noise fluctuations \reef{eq.noisyBoundary} also imply a fluctuation in the current $\delta j_i$, such that \reef{eq.noiseCurrent} is satisfied.
 
 Note in particular that \reef{eq.noisyBoundary} implies that the phase is a random variable. In our numerics we find that it is sufficient to update the noise boundary condition at larger intervals, say every 100 time steps. We have also run the simulations updating the noise at every time step, as well as every $10$ time steps, again with no apparent impact on the results - provided the noise amplitude is adjusted in accordance with \reef{eq.noisyBoundary}. Clearly there is a lower limit on the sampling frequency for which this statement is true - consider, {\it e.g.},  the extreme case of only updating the noise once or twice during the entire simulation. However our results  show no detectable dependence on sampling frequency, which implies that we stayed far away from this lower limit throughout. An illustration of the dependence of the evolution on the noise parameter is given in Supplementary Fig. \ref{fig.bulkNoise}.
 
\subsection{Quasinormal Mode Analysis}\label{app.QNM}
We find it convenient to perform this analysis using Schwarzschild coordinates, in which the background metric takes the form
\be\label{eq.SchwarzschildForm}
ds^2 = \frac{\ell^2}{z^2} \left[ -f(z)d\tau^2 + \frac{dz^2}{f(z)} + dx^2 \right]\,,
\ee
with the same function  $f = 1-z^2/z_h^2$ used throughout and Schwarzschild time $\tau$. We then expand the full equations of motion to linear order around a background solution. An important simplification is that in order to determine poles in the correlation function of the order parameter we only need to consider scalar operators, since these cannot mix with vector modes to first order in perturbation theory. \\
We start with an analysis of the normal phase.
Infinitesimal gauge transformations of the background configuration induce first-order changes in the metric scalar and gauge fields. In order to extract physical poles of correlation functions it is most convenient to determine the minimal set of gauge-invariant fluctuations containing the mode of interest (here the order-parameter fluctuation). In the broken-symmetry phase this requires more work, but in the symmetric phase the analysis is very simple. A suitable set of gauge-invariant modes are given by the real and the imaginary part of the scalar field fluctuation around zero background values
\be
a(z,\tau,x) =  e^{-i\omega \tau + i kx}\alpha(z)\,,\qquad b(z,\tau,x) = e^{-i\omega \tau + i kx}\beta(z)\,.
\ee
The modes $\alpha$ and $\beta$ decouple from all other fluctuations. This decoupling even happens when one goes beyond the probe limit and allows fluctuations of the metric, see, {\it e.g.}~\cite{Bhaseen:2012gg}. A more detailed discussion of the gauge-invariant modes for the probe system is presented below in the context of the broken phase. 
The modes we just identified satisfy the equations
\bea\label{eq.NormalPert}
\alpha'' + \left( \frac{f'}{f} - \frac{1}{z} \right)\alpha' + \frac{1}{f^2}\left( \omega^2 - fk^2 + T_0^2 \right)\alpha + \frac{2i\omega T_0}{f^2}\beta &=& 0\,,\nn
\beta'' + \left( \frac{f'}{f} - \frac{1}{z} \right)\beta' + \frac{1}{f^2}\left(  \omega^2 - fk^2 + T_0^2  \right)\beta - \frac{2i\omega T_0}{f^2}\alpha &=& 0\,.
\eea
Since the equations are linear in $\omega$, but quadratic in $k$, these admit an expansion \cite{Maeda:2009wv} in small frequency and momentum
\bea
\alpha &=& \alpha_{(0,0)}(z) + \omega\alpha_{(1,0)}(z) + k^2 \alpha_{(0,2)}(z) + \cdots\,,\nn
\beta &=& \beta_{(0,0)}(z) + \omega\beta_{(1,0)}(z) + k^2 \beta_{(0,2)}(z) + \cdots\,,
\eea
where
\be
\alpha_{(i,j)} \sim \alpha^s_{(i,j)} + z^2 \alpha^v_{(i,j)}+ \cdots \qquad \beta_{(i,j)} \sim \beta^s_{(i,j)} + z^2 \beta^v_{(i,j)}\,,
\ee
for $z\rightarrow z_{\rm UV}$. In this expression a superscript $s$ denotes source behavior, while $v$ denotes expectation value.
Thus Green functions of the dual operator have the small $\omega, k$ behavior
\be\label{eq.Green}
G(\omega, k) = \frac{\alpha_{(0,0)}^v + \omega \alpha_{(1,0)}^v + k^2 \alpha_{(0,2)}^v}{\alpha_{(0,0)}^s + \omega \alpha_{(1,0)}^s + k^2 \alpha_{(0,2)}^s}:=\frac{Z(\omega,k)}{ic \omega + k^2 + 1/\xi^2}\,,
\ee
as claimed in \reef{eq.DynamicResponse}
 ~above. More precisely, the Green functions of the operators dual to $\alpha$ and $\beta$ mix, but the eigenvalues of the matrix of correlation functions will be of the functional form \reef{eq.Green}. This establishes analytically that the dynamical critical exponent is $z=2$. In order to determine $\xi$, and thus $\xi_0$ and $\nu$, we solve the system of equations \reef{eq.NormalPert} numerically, using a finite difference discretization. Results of this analysis are described in more detail in Sec. II.
  ~above, and summarized in Fig. \ref{fig.UniversalityClass}.\\

 \label{sec.BrokenQNM}
We now describe the more involved calculation in the broken phase, where for $\omega\neq 0$ and $k\neq 0$ more mode mixing occurs. If we were to go beyond the probe approximation, the sound channel \cite{Policastro:2002tn}, holographically encoded as a scalar fluctuation of the metric, will also contribute. This means that in general the condensate-condensate two points function below $T_c$ no longer takes the simple form \reef{eq.DynamicResponse}.

Below the critical temperature, i.e., when the background contains a nontrivial scalar field,  we have to take into account the modes $e^{-i\omega \tau + i k x}\left\{ a_\tau(z), a_z(z), a_x(z), \alpha(z),\beta(z)\right\}$, where $a_\mu(z)$ are perturbations of the gauge field and $\alpha(z)$ and $\beta(z)$ are perturbations of the real and imaginary parts of the complex scalar field respectively. Not all of these modes are physical, since we can generate an infinitesimal perturbation by acting on the background with a gauge transformation $e^{-i\omega \tau + i kx}\lambda(z)$. This generates the perturbation modes $\left\{ \delta a_\tau,\delta a_x, \delta a_z \right\}=e^{-i\omega \tau + i k x}\{ -i\omega \lambda(z), -ik \lambda (z),\lambda'(z)  \}$ in the gauge sector, as well as a perturbation of the imaginary part of the scalar $\delta\beta = e^{-i\omega \tau + i k x} a_0(z) \lambda$. We cannot generate a real part of the scalar perturbation in this way, because the background value $a_0(z)$ is purely real. Thus we have three gauge-invariant perturbations
\be
\Phi_1 = \alpha\,,\qquad \Phi_2 = i\omega \beta + a_0 a_\tau\,,\qquad \Phi_3=-ik \beta + a_0 a_x\,.
\ee
The last mode, $\Phi_3$ transforms as a vector, but the combination $k\Phi_3$ is a scalar, so that it too can contribute to the linear equations for the condensate fluctuation. Since this combination vanishes at zero momentum, it does not appear in the dynamic susceptibility calculation.
\subsubsection*{Dynamic Susceptibility: $k=0$}
In this case the vector like perturbation $\Phi_3$ decouples, and we have the equations
\bea\label{eq.BrokenPert}\Phi_1'' + \left( \frac{f'}{f} - \frac{1}{z} \right)\Phi_1' + W_{11}[T_0] \Phi_1 + W_{12}[T_0,a_0,\omega]\Phi_2 &=& 0\,,\nn
\Phi_2'' + V_2[T_0,a_0,\omega]\Phi_2' +  W_{22}[T_0]  \Phi_2 + W_{21}[T_0,a_0,\omega]\Phi_1 &=& 0\,.
\eea
with the coefficients
\bea
W_{11} &=& \frac{\omega^2 + T_0^2}{f^2}\,,\qquad W_{12} = \frac{2i\omega T_0}{f^2}\,, \qquad W_{21} = -\frac{2iT_0}{\omega z^2 f^2}\left(  \omega^2 z^2 - 2 a_0^2 f   \right)\,,\nn
V_2&=&\frac{\omega^2 z^2 (zf' -f) - 2a_0f^2(a_0-2za_0')}{zf(\omega^2z^2-2a_0^2f)}\,,\nn
W_{22} &=& \frac{1}{z^2 f^2 \left( \omega^2 z^2 - 2 a_0^2 f  \right)} \Bigl[  4 a_0^4 f^2 + \omega^2 z^4 \left( \omega^2 + T_0^2 \right) - 2 z^2 a_0^2 f \left(  2\omega^2 + T_0^2  \right) - 4z^2 f^3 a_0'^2 \Bigr.\nn
&& \Bigl. - 2z a_0 a_0'f^2 \left(  zf' - 2 f  \right) \Bigr]\,.
\eea
For completeness we also present the decoupled equation for the remaining gauge-invariant mode
\be
\Phi_3'' + \left(   \frac{1}{z} + \frac{f'}{f} \right)\Phi_3' + \frac{\omega^2 z^2 - 2 a_0^2 f}{z^2 f^2}\Phi_3=0\,.
\ee
The behavior of these poles in the complex plane is summarized in Supplementary Fig. \ref{fig.dynamicSusceptBroken}.
\subsubsection*{Static Susceptibility: $\omega=0$}
For the computation of the static susceptibility, $\Phi_2$ decouples from the other two modes, which satisfy the coupled equations 

\bea\label{eq.BrokenPertStat}\Phi_1'' + \left( \frac{f'}{f} - \frac{1}{z} \right)\Phi_1' - \frac{k^2 f - T_0^2}{f^2}\Phi_1 + \frac{2 a_0T_0}{f^2}\Phi_3 &=& 0\,,\nn
\Phi_3'' + \frac{1}{z}\Phi_3' - \frac{k^2 z^2 + 2 a_0^2}{z^2 f}\Phi_3 - \frac{4 a_0 T_0}{z^2 f}\Phi_1 &=& 0\,.
\eea
The remaining equation for the mode $\Phi_2$ reads
\be
\Phi_2'' + \left(   \frac{f'}{f}  - \frac{1}{z} + \frac{4 a_0 \left(a_0 -  z a_0'  \right) }{z\left(k^2 z^2 + 2 a_0^2  \right)}\right)\Phi_2' + \left( \frac{4a_0'(za_0' - a_0)}{z(k^2 z^2 + 2 a_0^2)} + \frac{T_0^2}{f^2} - \frac{z^2k^2 + 2 a_0^2}{z^2 f} \right) \Phi_2 = 0\,.
\ee
The behavior of these poles in the complex plane is summarized in Supplementary Fig. \ref{fig.staticSusceptBroken}.



\vspace{0.21cm}

\noindent
{\bf Author Contribution:}\\
\noindent
The authors jointly defined the project. JS developed and carried out the numerical simulations. All authors contributed to the analysis and interpretation of the  numerical data and the preparation of the manuscript.

\vspace{0.21cm}

\noindent
{\bf Additional Information}\\
\noindent
{\it Competing Financial Interests:}
The authors declare no competing financial interests.
\renewcommand{\figurename}{Supplementary FIG.}
\setcounter{figure}{0}
\appendix
\newpage
\section{Supplementary Material}
 \begin{figure}[h!]
\begin{center}
\textsf{a}\includegraphics[width=0.45\textwidth]{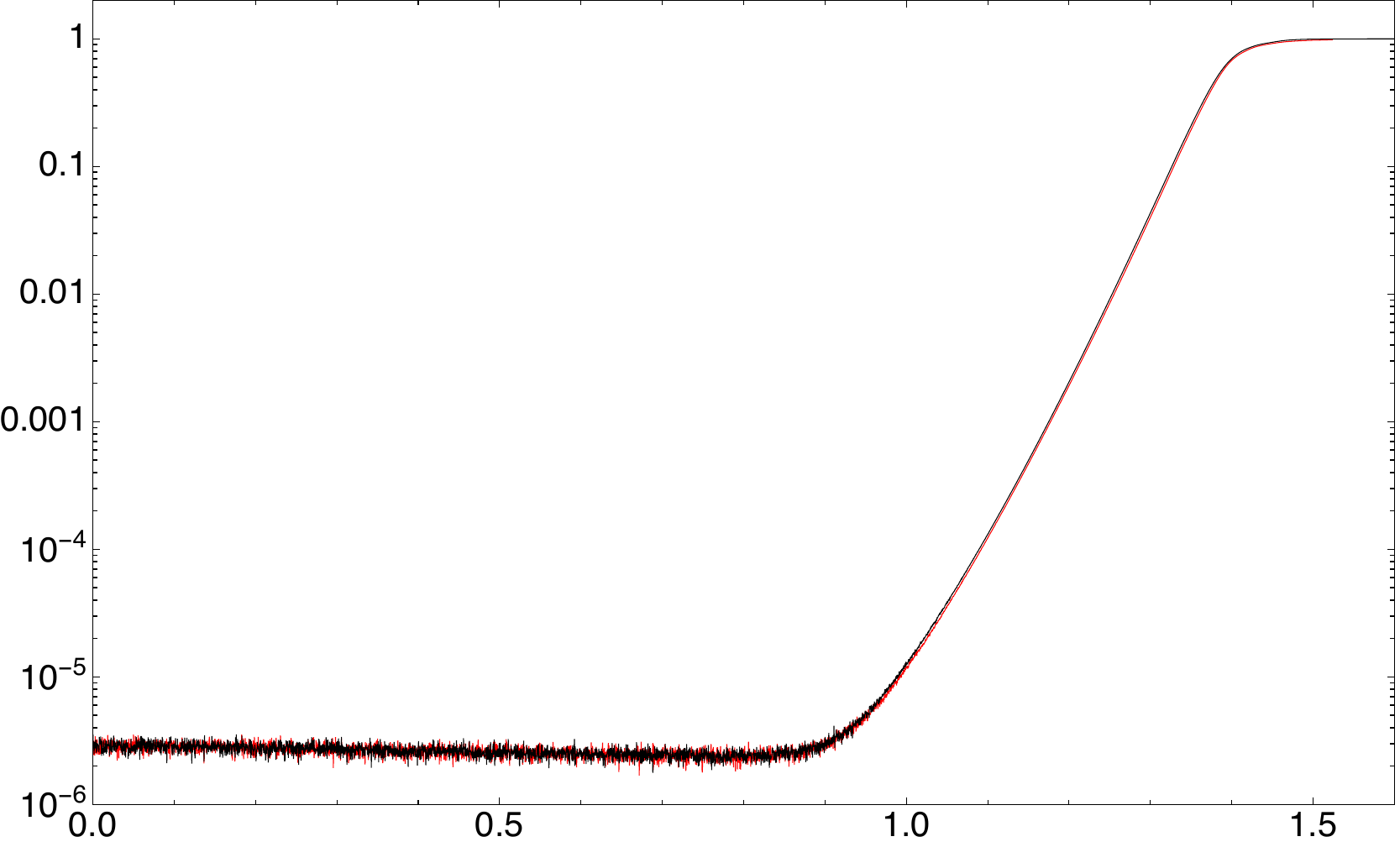}\put(-77,27){\makebox(0,0){\begin{sideways}{\small $\langle {\cal O}(t)\rangle/\langle {\cal O}(\infty)\rangle$}\end{sideways}}}\put(-34,-2){\makebox(0,0){$t/\tau_{\rm Q}$}}\hskip3em\textsf{b}\includegraphics[width=0.45\textwidth]{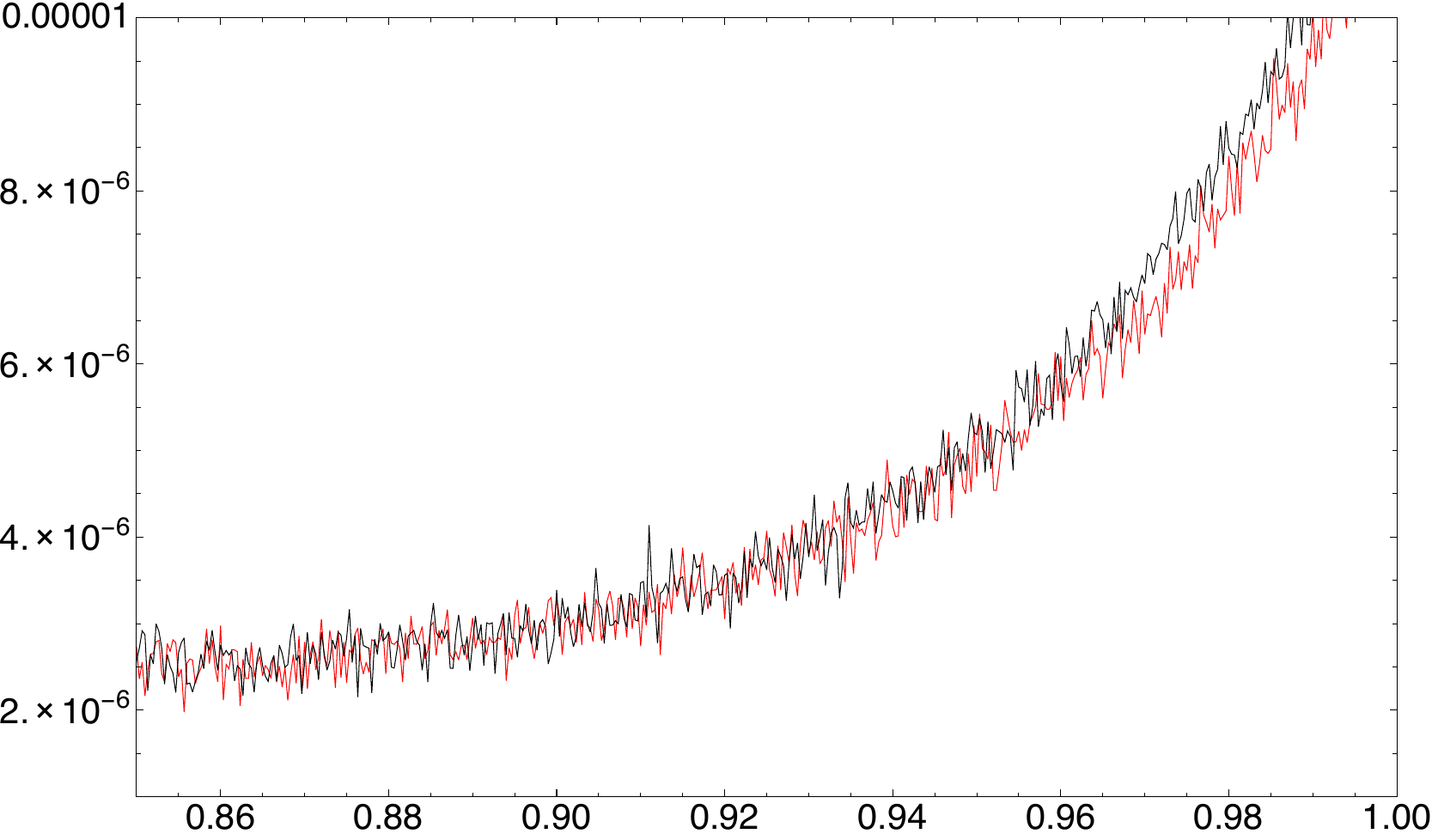}\put(-77,27){\makebox(0,0){\begin{sideways}{\small $\langle {\cal O}(t)\rangle/\langle {\cal O}(\infty)\rangle$}\end{sideways}}}\put(-34,-2){\makebox(0,0){$t/\tau_{\rm Q}$}}
\caption{ {\bf Noise dependence of the order parameter dynamics.} Panel a: Average over the ring of condensate density as a function of time for two different values of the noise amplitude differing by a factor of ten, shown in red and black. Panel b shows a blowup of the `knee' region where the condensation process first happens. We see that this process is largely insensitive to the noise amplitude. \label{fig.bulkNoise}}
\end{center}\end{figure}
\clearpage
\begin{figure}[h!]
\begin{center}
\textsf{a}\includegraphics[width=0.4\textwidth]{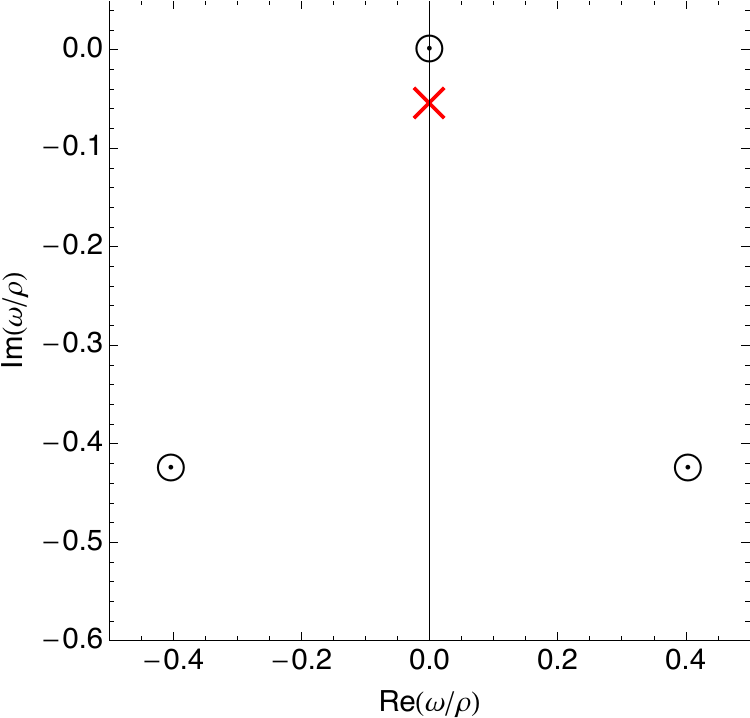}\hskip4em\textsf{b}\includegraphics[width=0.4\textwidth]{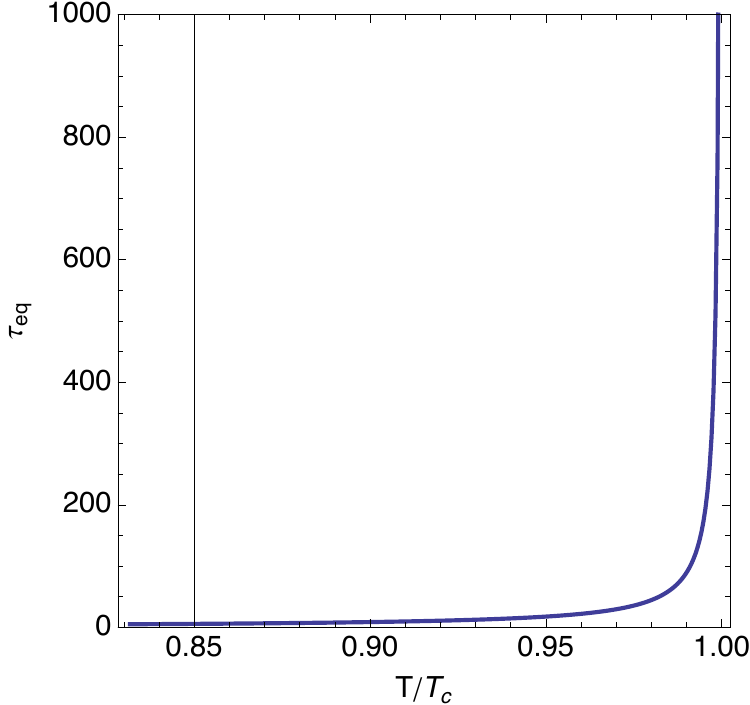}
\caption{{\bf Dynamic susceptibility in the broken phase.} Panel \textsf{a}: Poles in dynamic correlation function at $T/T_c=0.9$. There is a Goldstone pole at the origin. The leading pole nearest the real axis, corresponding to the amplitude or `Higgs' mode, with imaginary part $\omega_\star$ gives the equilibration time. Panel \textsf{b}: inverse of the imaginary part of leading pole $\omega_\star$ as the critical point is approached. This directly gives the equilibration time $\tau$. The best fit result gives $\tau \sim 0.8 \epsilon^{-1}$, with a $\tau_0$ differing from the unbroken phase. Again the explicit calculation agrees with the analytical derivation that $z=2$. The pole structure found here resembles closely the AdS$_4$ results of \cite{Amado:2009ts,Bhaseen:2012gg}.\label{fig.dynamicSusceptBroken}}
\end{center}\end{figure}

\clearpage
\begin{figure}[h!]
\begin{center}
\textsf{a}\includegraphics[width=0.4\textwidth]{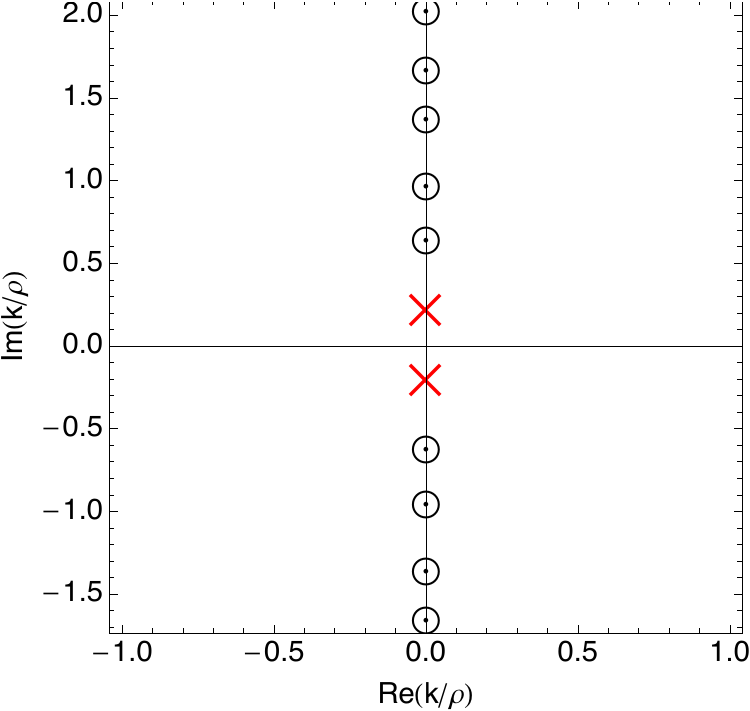}\hskip4em\textsf{b}\includegraphics[width=0.4\textwidth]{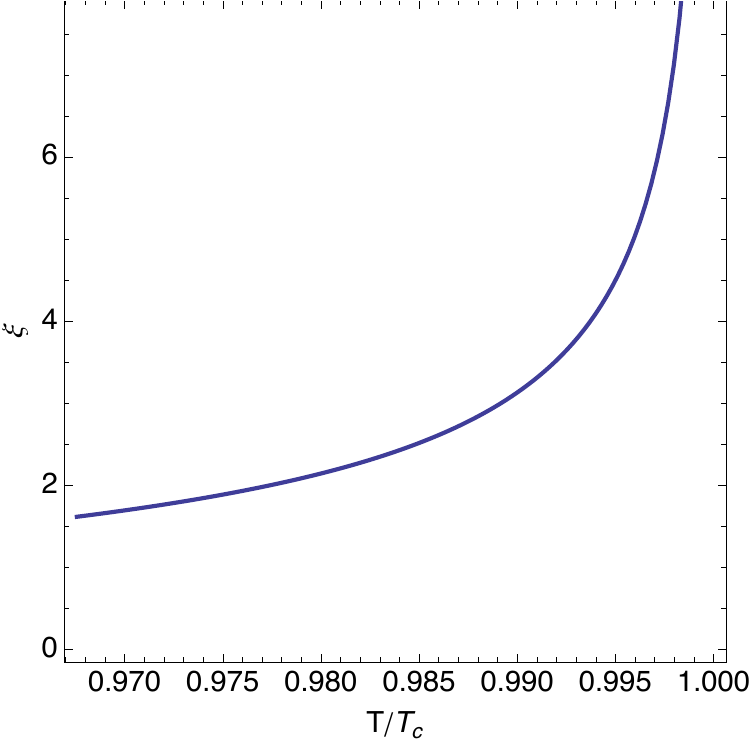}
\caption{ {\bf Static susceptibility in broken phase.} Panel \textsf{a}: poles in static correlation function at $T/T_c=0.9$. The leading poles nearest the real axis, with imaginary part $k_\star$ give the correlation length. Panel \textsf{b}: imaginary part of leading pole $k_\star$ as the critical point is approached from below. The correlation length $\xi = k_\star^{-1}$ diverges as $\xi = \xi_0\epsilon^{-1/2}$. We determined $\xi_0 = 0.39 \pm 0.01$ from numerically solving Eq. \reef{eq.NormalPert}. \label{fig.staticSusceptBroken}}
\end{center}\end{figure}

%


\end{document}